\documentstyle[epsfig]{ioplppt}       
                                         
\def\lapproxeq{\lower .7ex\hbox{$\;\stackrel{\textstyle                                         
<}{\sim}\;$}}                                         
\def\gapproxeq{\lower .7ex\hbox{$\;\stackrel{\textstyle                                         
>}{\sim}\;$}}                                         
       
\newcommand{\be}{\begin{equation}}       
\newcommand{\ee}{\end{equation}}       
\newcommand{\beqn}{\begin{eqnarray}}       
\newcommand{\eeqn}{\end{eqnarray}}

\newcommand{\fP}{I\!\!P}       
\def\funp{{I\!\!P}}      
\begin{document}                                         
       
\begin{flushright}
DTP/99/78 \\ 
September 1999 \\
(revised October 1999)
\end{flushright}\vspace*{-0.5cm}       
       
\title{The QCD description of diffractive processes}       
\jl{4}       
\vspace{0.5cm}       
\author{M. W\"usthoff and A.D. Martin}        
       
\address{University of Durham, Department of Physics,       
Durham DH1 3LE, UK}       
       
\begin{abstract}  
We review the application of perturbative QCD to diffractive processes.  We introduce the   
two gluon exchange model to describe diffractive $q\bar{q}$ and $q\bar{q}g$ production in   
deep inelastic scattering.  We study the triple Regge limit and briefly consider multiple gluon   
exchange.  We discuss diffractive vector meson production at HERA both at $t = 0$ and large   
$|t|$.  We demonstrate the non-factorization of diffractive processes at hadron colliders.   
\end{abstract}       
       
\section{Introduction}       
Diffractive scattering        
has become a vast area of study in particle physics and generated a wide range of       
theoretical approaches.  Here we will concentrate on those processes       
which have the potential to provide a deeper insight into the dynamics       
of strong interactions, that is into quantum chromodynamics (QCD). Our main focus will be        
on diffractive deep inelastic scattering. Diffraction in $p\bar{p}$-scattering        
or photoproduction will be discussed fairly briefly.        
The motivation for studying       
diffractive deep inelastic scattering is not only driven by the fact that       
we have excellent data, but also that it offers an opportunity to explore the interesting        
transition from \lq\lq hard\rq\rq~to \lq\lq soft\rq\rq~physics.        
Hard physics is associated with the  well established        
parton picture and perturbative QCD, and is applicable to processes for which a large scale is        
present.       
Soft dynamics on the other hand, associated for example with total cross sections and elastic        
scattering of hadrons, is described by non-perturbative aspects of QCD. Diffractive deep        
inelastic scattering offers the opportunity to directly       
probe the semihard transition region and to link the, otherwise distinct, regimes of soft and        
hard physics. The means by which       
we attack the problem is perturbative QCD carefully       
extrapolated into the semihard regime. It is thus natural to       
put strong emphasis on diffractive processes, which are believed to be purely       
perturbative, such as        
diffractive vector meson production in deep inelastic       
scattering or in photoproduction when the momentum transfer is large.       
       
The classical       
definition of diffraction in hadron-hadron or (virtual) photon-hadron scattering is the quasi        
elastic       
scattering of one hadron combined with the dissociation of the second hadron or 
photon\footnote{We consider here only the diffractive dissociation of one of the incoming 
particles.}.  For        
example, at the electron-proton collider HERA, diffractive deep inelastic scattering is usually        
written       
\be       
\gamma^* + p \; \rightarrow \; X + Y       
\label{eq:a1}       
\ee       
where $Y$ denotes the elastically scattered proton (or one of its low mass excitations) and        
$X$ denotes all the hadrons coming from the dissociating virtual photon.       
The classical method of detection is to tag the       
elastically scattered hadron. In the present high energy collider experiments, however,       
the elastically scattered protons usually disappear into the beampipe       
requiring special detectors far down, and adjacent to, the beampipe for their detection.       
Very often these detectors cannot deliver high enough statistics       
or good enough energy resolution so that an alternative method, the rapidity       
gap method, is employed. The rapidity gap method is based on the simple        
fact that the elastically scattered proton (or $Y$ state) travels in roughly the same direction as        
the original hadron leaving a large gap between the rapidity of this particle and the other       
outgoing hadrons forming the inclusive state $X$.  Rapidity is a measure of the component of       
a particle's velocity along the incoming proton direction. That is diffraction       
at the HERA electron-proton collider is characterized by       
the quasi-elastic scattering of the proton and the dissociation       
of the virtual photon.  Such a diffractive event is in contrast to the usual deep inelastic       
scattering event       
\be       
\gamma^* + p \; \rightarrow \; X       
\label{eq:a2}       
\ee       
in which the virtual photon breaks up the proton and the outgoing hadrons populate the full        
interval of rapidity.  Nevertheless about 10\% of all events in deep inelastic scattering       
are of diffractive nature and high statistics data samples have been       
collected at HERA \cite{HERAd}.        
       
Before this diffraction was observed       
in hadron collisions and is still being measured at the Tevatron proton-antiproton collider at        
the present time.  The relative rate of diffraction compared to the total rate is on a 
similar level to that of deep inelastic        
scattering \cite{SOFTTev}. The problem, however, is that the total       
cross section is huge       
and only a tiny fraction of events are of special interest such as        
heavy particles ($W$ and $Z$ bosons or the top quark) and jets of high transverse  
momentum.  Since the hadronic activity is much higher in hadron-hadron scattering       
compared to deep inelastic scattering        
the detection of rapidity gaps is a       
tedious task. With a trigger set on $W$ bosons or high-$p_T$ jets       
the fraction of events with a rapidity gap is only on the       
level of 1\% \cite{Tev}. It requires a special effort to extract diffractive       
processes at the Tevatron whereas at HERA they almost come for free.       
       
There is a group of processes which do not precisely belong       
to diffraction in the way we have defined it.  These are processes with       
large momentum transfer across the rapidity gap.  Examples are the production of opposite       
di-jets with a rapidity gap in between them at the Tevatron, or diffractive vector       
meson production at HERA. Due to the large momentum transfer the proton       
(or antiproton) always breaks up. Because of the large rapidity gap,        
however, they are closely related to the traditional diffractive processes.       
\begin{figure}[htb]        
  \vspace*{0cm}       
\begin{center}      
         \epsfig{figure=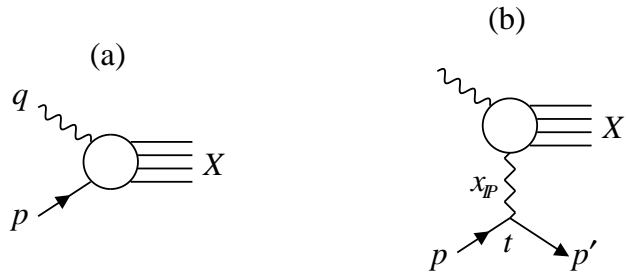,height=5.5cm}       
\end{center}       
\vspace*{-1cm}       
\caption{\it Diagrams for DIS, $\gamma^* p \rightarrow X$, and inclusive diffractive DIS,       
$\gamma^* p \rightarrow Xp$.  In the latter process there is a rapidity gap between the       
outgoing hadrons $X$ and the slightly deflected proton.  In the diagram the gap has been       
associated with Pomeron exchange carrying a fraction $x_\funp$ of the proton's momentum. 
\label{diag1_a}}       
\end{figure}       
       
The early attempts to describe diffractive scattering were based on       
Regge phenomenology where the Pomeron is thought to be the leading        
Regge pole with a well defined and unique, process-independent        
Regge trajectory. The Pomeron trajectory, together with        
secondary trajectories, allows one       
to fit all kinds of hadronic data (total and elastic cross sections)       
\cite{DL}. These universal trajectories $\alpha_R (t)$ form the backbone of Regge       
phenomenology.  According to early analyses it seemed that diffraction can be described in  
terms of triple Regge diagrams (see Section 3) without changing the trajectories or       
the couplings to hadrons. The only free parameters were the couplings       
of the Reggeons amongst themselves.  The more recent results \cite{SOFTTev} from the Tevatron,       
however, indicate that this simple picture does not hold at very high energies.       
The extrapolation of the cross section from the Sp$\bar{\rm p}$S CERN collider energy to        
the Tevatron energy clearly overshoots the data. The explanation for the discrepancy       
is most likely given by unitarity corrections.        
 
After introducing the structure functions that can be measured in diffractive deep inelastic  
scattering, we outline in Section 3 the Regge description of diffractive processes based on the  
Pomeron, or the \lq\lq soft\rq\rq~Pomeron as it is now frequently called.  The perturbative  
QCD description of diffractive processes, which is the subject of this review, is based on two  
gluon exchange.  This will form the framework of our discussions from Section 4 onwards.   
Two gluon exchange is the simplest form of vacuum quantum exchange and, for historical  
reasons, people often speak of the \lq\lq hard\rq\rq~Pomeron.  However the connection  
between \lq\lq soft\rq\rq~(non-perturbative) and \lq\lq hard\rq\rq~(perturbative) diffraction is  
far from clear.  The different energy behaviours are very evident, as will be seen, for example,  
when we discuss diffractive vector meson photo- and electro-production (see, in particular,  
Fig.~9). 
       
\section{Diffractive structure functions}      
      
First let us recall (inclusive) deep inelastic electron-proton scattering, $ep \rightarrow eX$,       
where $X$ represents all the fragments of the proton which has been broken up by the high       
energy electron.  The basic subprocess $\gamma^* p \rightarrow X$ is shown in Fig.~1(a).  It       
can be expressed in terms of two functions $F_2$ and $F_L$ which characterize the structure       
of the proton.  These proton structure functions depend on two (invariant) variables, the \lq\lq       
virtuality\rq\rq~of the photon $Q^2 \equiv -q^2$ and the Bjorken $x$-variable      
\be      
\label{eq:a3}      
x \; \equiv \; \frac{Q^2}{2 p.q} \; = \; \frac{Q^2}{Q^2 + W^2},      
\ee      
where $p$ and $q$ are the four-momenta of the proton and virtual photon respectively, see       
Fig.~1(a).  $W$ is the total $\gamma^* p$ centre-of-mass energy. If we were to view the     
proton as made up of massless, point-like quark constituents, then it is easy to show that $x$       
is the fraction of the proton's momentum carried by the quark struck by the virtual photon.  In     
this simple quark model $F_L = 0$ and    
\be    
\label{eq:b3}    
F_2 \; = \; F_T \; = \; \sum_q \: e_q^2 \: xq (x)    
\ee    
is independent of $Q^2$.  The sum is over the flavours of quarks, with electric charge $e_q$     
(in units of $e$) and distributions $q (x)$.  $F_{T,L}$ are the proton structure functions for     
DIS by transversely, longitudinally polarised photons.      
      
The general form of the DIS cross section, up to target mass corrections, is      
\be      
\label{eq:a4}      
\frac{d^2 \sigma (ep \rightarrow eX)}{dxdQ^2} \; = \; \frac{2 \pi \alpha^2}{x Q^4} \: \left \{       
[1 + (1 - y)^2] \: F_2 (x, Q^2) \: - \: y^2 F_L (x, Q^2) \right \}      
\ee      
where $\alpha$ is the electromagnetic coupling.  The third variable $y$ is needed to fully       
characterize the DIS process, $ep \rightarrow eX$, namely $y = Q^2/xs$ where       
$\sqrt{s}$ is the total       
centre-of-mass energy of the electron-proton collision.      
      
Now we turn to inclusive diffractive DIS, $ep \rightarrow eXp$, for which the subprocess is       
$\gamma^* p \rightarrow Xp$, where a slightly deflected proton and the cluster $X$ of       
outgoing hadrons are well separated in rapidity.  In terms of Regge terminology the rapidity       
gap is associated with Pomeron (or vacuum quantum number) exchange, shown by the       
zig-zag line in Fig.~1(b).  The $\gamma^* p \rightarrow Xp$ subprocess is described by       
diffractive structure functions $F_{2,L}^D$, which now depend on four variables: $\beta       
\equiv x/x_\funp$, $Q^2$, $x_\funp$ and $t$.  The variable $x_\funp$ is the fraction of the       
proton's momentum that is carried away by the Pomeron.  The variable       
$\beta$ in diffractive DIS plays an       
analogous role to $x$ in DIS      
\be      
\label{eq:a5}      
\beta \; \equiv \; \frac{x}{x_\funp} \; = \; \frac{Q^2}{2 x_\funp p.q} \; = \; \frac{Q^2}{Q^2 +       
M^2}      
\ee      
where $M$ is the invariant mass of the diffractively produced $X$ cluster in Fig.~1(b).      
      
In analogy to (\ref{eq:a4}), the general form of the diffractive DIS cross       
section is      
\begin{eqnarray}      
\frac{d^4 \sigma^D (ep \rightarrow eXp)}{d \beta d Q^2 dx_\funp dt} & = & \frac{2 \pi       
\alpha^2}{\beta Q^4} \left \{ [1 + (1 - y)^2] \: F_2^{D (4)} (\beta, Q^2, x_\funp, t) \right .       
\nonumber \\      
& & \nonumber \\      
& & -y^2 \left . F_L^{D (4)} (\beta, Q^2, x_\funp, t) \right \}.      
\label{eq:a6}      
\end{eqnarray}      
The superscript (4) is to indicate that the structure functions depend on four independent       
variables.  The $F_L$ contribution is often neglected due to the smallness of $y^2$ in most of       
the measurements.  The variable $t = (p - p^\prime)^2$ is the square of the 4-momentum       
carried by the Pomeron.  Most diffractive DIS events occur for small values of $t$.  Usually       
$t$ is integrated over and measurements of the structure function      
\be      
F_2^{D (3)} (\beta, Q^2, x_\funp) \; = \; \int \: dt \: F_2^{D (4)} (\beta, Q^2, x_\funp, t)      
\label{eq:a7}      
\ee      
are made.      
      
\section{Regge approach to diffraction}      
      
The early attempts to describe diffractive scattering were based on Regge theory   
(see, for example, \cite{PDBC}), in which the basic idea is that sequences of hadrons of mass 
$m_i$ and spin   
$j_i$ lie on Regge trajectories $\alpha (t)$ such that $\alpha (m_i^2) = j_i$.  Prior to QCD,   
strong interactions were thought to be due to the exchange of complete trajectories of   
particles.  Indeed the Regge model is able to successfully describe all kinds of \lq\lq   
soft\rq\rq~high energy hadronic scattering data:  differential, elastic and total cross section   
measurements.  In this model the high energy behaviour of a hadron scattering amplitude at   
small angles has the form      
\be      
\label{eq:a8}      
A (s, t) \; \sim \; \sum_R \: \beta (t) \: s^{\alpha_R (t)}      
\ee      
where for simplicity we have omitted the signature factor.  $s$ is the square of the       
centre-of-mass energy and $-t$ is the square of the four-momentum transfer.  The observed  
hadrons were found to lie on trajectories $\alpha_R (t)$ which are approximately linear in $t$  
and parallel to each other.  That is hadrons of increasing spin and mass, but with the other  
quantum numbers the same, lie on a single trajectory $\alpha_R (t)$.  The leading such  
trajectories are the $\rho, a_2, \omega$ and $f$ trajectories which are all approximately  
degenerate with 
\be      
\label{eq:a10}      
\alpha_R (t) \; \simeq \; 0.5 \: + \: 0.9 t.      
\ee 
For example only the $\rho$ trajectory has the appropriate quantum numbers to be exchanged  
in the process $\pi^- p \rightarrow \pi^0 n$.  The energy or $s$ dependence of the differential  
cross section $d\sigma/dt$ therefore determines $\alpha_\rho (t)$ for $t < 0$, see  
(\ref{eq:a8}).  For small $|t|$ the trajectory $\alpha_\rho (t)$ is found to be linear in $t$ and,  
when extrapolated to positive $t$, to pass through the $\rho (1^-)$ and $\rho (3^-) \ldots$  
states, i.e. $\alpha_\rho (m_\rho^2) = 1, 3, \ldots$ at the appropriate mass values. 
 
So far so good, but the total cross sections are observed to increase slowly with energy at high  
energies, so we need a higher lying trajectory.  This is best seen from the optical theorem  
which expresses a total cross section (say, for $AB \rightarrow X$) in terms of the imaginary  
part of the forward ($AB \rightarrow AB$) elastic scattering amplitude 
\be 
\label{eq:b10} 
\sigma (AB \rightarrow X) \; = \; \frac{1}{s} {\rm Im} A (s, 0) \; = \; \sum_R \:  
\beta_R \: s^{\alpha_R (0) - 1}. 
\ee 
To account for the asymptotic energy dependence of the total cross sections a  
Pomeron\footnote{Originally the total cross sections were thought to asymptote to a constant  
at high energies and so a Pomeron with $\alpha_\funp (0) = 1$ was introduced \cite{POM}.}  
(or vacuum quantum number exchange) trajectory is invoked with intercept $\alpha_\funp (0)  
\sim 1.08$.  Indeed the total, elastic and differential hadronic cross section data are found to  
be well described (for small $|t|$) by taking a universal pole form for the Pomeron 
\be      
\label{eq:a9}      
\alpha_\funp (t) \; \simeq \; 1.08 \: + \: 0.25 t,      
\ee      
together with the other sub-leading trajectories as in (\ref{eq:a10}) \cite{DL}.  The Pomeron  
should be regarded as an effective trajectory, since the $s^{0.08}$ power behaviour of the  
total cross sections will ultimately violate the Froissart bound.  The link between this  
successful Regge description of \lq\lq soft\rq\rq~processes and       
the underlying fundamental theory of QCD is not, as yet, known in detail.  Most probably 
Pomeron exchange originates mainly from the exchange of a two-gluon bound state, while the 
meson trajectories $(\rho, a_2, \omega, f)$ correspond to $q\bar{q}$ bound states.  The 
Regge Pomeron discussed above is now often called the \lq\lq soft\rq\rq~Pomeron. 
\begin{figure}[htb]        
  \vspace*{0cm}       
     \centerline{       
         \epsfig{figure=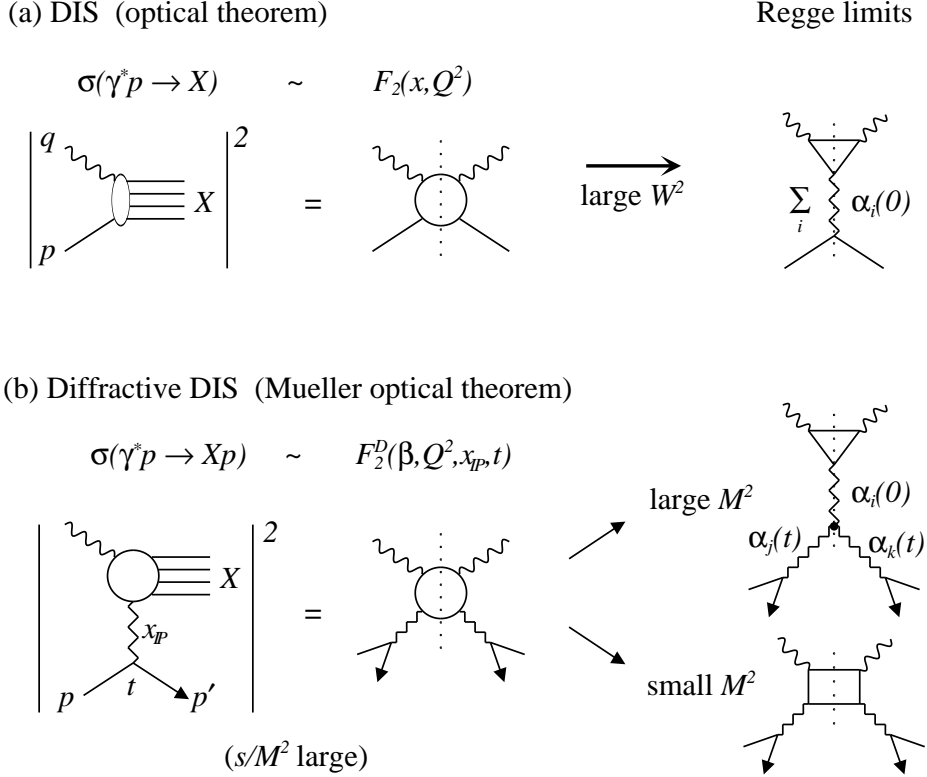,width=15cm}      
           }       
\vspace*{-2cm}       
\caption{\it The DIS and diffractive DIS cross sections expressed symbolically as structure       
functions via the optical theorem and its generalisation.  The Regge limits are shown in the       
rightmost diagrams (where the photon couples via a quark configuration).  A summation over       
allowed Regge exchanges is implied in (b).  
 \label{diag1_b}}       
\end{figure}       
      
To apply this approach to inclusive DIS and, more especially to its diffractive       
component, we again make use of the optical theorem, together with its generalisation by  
Mueller \cite{MOT},       
which are symbolically shown by the equalities in Fig.~2.  The optical theorems       
express the total cross sections in terms of the imaginary parts of the 2-body (or 3-body)  
forward elastic       
scattering amplitudes, or to be precise the discontinuities of the amplitudes across       
the cuts along the $W^2$ (or $M^2$) axes, which are indicated by the dotted lines       
in Fig.~2(a) (or (b)).  The last diagrams show the various Regge limits for the structure       
functions, where the coupling to the photon is via a quark line.  For DIS this gives      
\be      
\label{eq:a11}      
F_2 \: \sim \sum_i \beta_i (W^2)^{\alpha_i (0) - 1} \quad \sim       
\quad \sum_i \beta_i x^{1 - \alpha_i (0)}      
\ee      
for small $x$, see (\ref{eq:a3}).  In the naive parton model the valence and       
sea quark contributions to $F_2$ are associated with meson and Pomeron exchange       
respectively, and so using (\ref{eq:b3}) we have      
\begin{eqnarray}      
\label{eq:a12}      
xq_V \: \sim \: x^{1 - \alpha_R (0)} \quad \sim \quad x^{0.5} \nonumber \\      
& & \nonumber \\      
x q_S \: \sim \: x^{1 - \alpha_\funp (0)} \quad \sim \quad x^{-0.08}      
\end{eqnarray}      
for small $x$.      
      
For diffractive DIS, $\gamma^* p \rightarrow X p$, we apply Mueller's generalisation of the       
optical theorem \cite{MOT}.  For this diffractive case, that is when $s/M^2$ is large, the theorem is       
shown pictorially by the first two diagrams of Fig.~2(b).  That is the cross section is given by       
the discontinuity across the $M^2$ cut of the (three-body) $\gamma^* p\bar{p}$ elastic       
amplitude, where a sum over the exchange Reggeons is implied.  The Regge prediction       
depends on whether $M^2$ is large or small.  For small $M^2$ the quark box gives the main       
contribution to photon-Pomeron scattering.      
In Ref.~\cite{DLdiff} a charge conjugation $C = 1$ vector current coupling of the Pomeron      
to quarks was       
introduced with a free coupling constant. The resulting contribution to diffraction    
due to the quark box diagram is found to be    
\be    \label{DL-eq}  
F_2^D \; \sim \; \beta (1 - \beta).      
\label{eq:a13}      
\ee      
This model shows already many of the properties which are also present in the two gluon     
exchange    
approach \cite{MD1} \footnote{The problem that this model breaks electromagnetic gauge invariance, and a possible     
remedy, has been discussed in Ref.~\cite{MD1}.}.     
It is important to note the leading  twist nature of     
diffractive deep inelastic scattering, that is $F_2^D$ is independent of $Q^2$. One also finds the strong alignment of the  
quark-antiquark pair along the photon-Pomeron axis   
which is a consequence of the low transverse momentum $k_t$ of the quarks relative to this  
axis.  The particular property of alignment has already been pointed out by Bjorken \cite{Bj}.    
It was argued in Ref.~\cite{DLdiff} that the quark-Pomeron coupling for inclusive DIS       
and diffractive DIS should be the same and by fitting inclusive data one can extract the       
coupling constant and make predictions for diffraction.  Indeed some 10\% of HERA DIS       
events were predicted to be diffractive.      
      
For large $M^2$, on the other hand, we have the double Regge limit ($s/M^2 \rightarrow       
\infty$ and $M^2 \rightarrow \infty$) and the diffractive structure function is described by a       
sum of triple Regge diagrams      
\be      
\label{eq:a14}      
F_2^D \: \sim \: \sum_{i,j,k} \beta_{ijk} \left ( \frac{s}{M^2} \right       
)^{\alpha_j (t) + \alpha_k (t)} \: (M^2)^{\alpha_i (0)}.      
\ee      
The leading behaviour, which is given by the triple Pomeron contribution, is      
\be      
\label{eq:a15}      
F_2^D \; \sim \; (M^2)^{\alpha_\funp (0) - 2 \alpha_\funp (t)} \; \sim \; 1/M^2.      
\ee      
      
\section{Two-gluon exchange model of diffraction}      
      
We now use perturbative QCD to describe the diffractive DIS process $\gamma^* p       
\rightarrow Xp$.  In QCD the \lq\lq Pomeron\rq\rq~(or vacuum quantum number exchange)       
is, in its simplest form, represented by two gluons \cite{LONU,GSLR}.  The minimum number of gluons to form a       
colourless state is of course two.  It is not excluded that more than two gluons are exchanged       
and it is important that whenever we talk about two-gluon exchange to remember there is the       
possibility to extend the formalism to multigluon exchange. One might object that the whole       
process is soft and perturbation theory not applicable.  
Saturation effects for high parton densities, however, screen soft contributions, so that      
a fairly large fraction of the cross section is hard and therefore       
eligible for a perturbative treatment \cite{Mue-sat}.

The two basic pQCD diagrams are shown in Fig.~3, in which the photon dissociates into       
either a $q\bar{q}$ pair or $q\bar{q}g$.  In the first diagram the $q$ and $\bar{q}$ carry       
fractions $\alpha$ and $(1 - \alpha)$ of the momentum of the photon, and have transverse       
momenta $\pm \mbox{\boldmath $k$}_t$.  In the second diagram these variables apply to the       
gluon and the $(q\bar{q})$ system.  At first sight it might appear that $q\bar{q}g$       
would be a small correction to $q\bar{q}$ production on account of an extra $\alpha_S$       
factor.  However $q\bar{q}g$ production dominates at large $M^2$.  It incorporates an extra       
$t$-channel spin 1 gluon in contrast to the (lower lying) $t$-channel spin $\frac{1}{2}$       
quark exchange of $q\bar{q}$ production.  In fact $q\bar{q}g$ is described by the triple       
Pomeron diagram and gives the leading contribution at       
large $M^2$ (or small $\beta$), whereas $q\bar{q}$ production is leading at small $M^2$,       
see Fig.~2(b).      
       
\begin{figure}[htb]        
  \vspace*{-1cm}       
     \begin{center}       
         \epsfig{figure=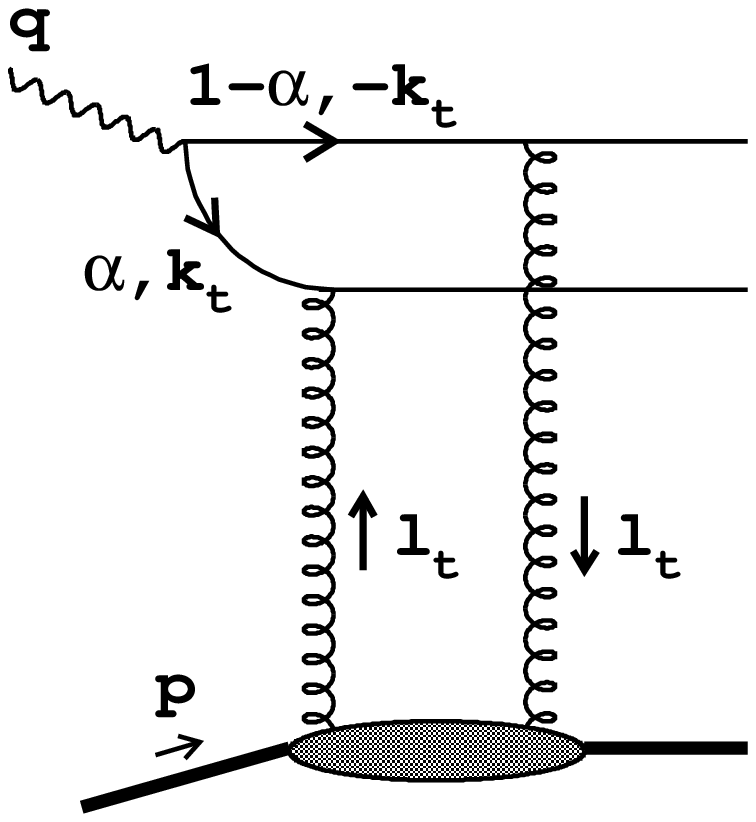,height=6cm}\hspace*{-1cm}       
         \epsfig{figure=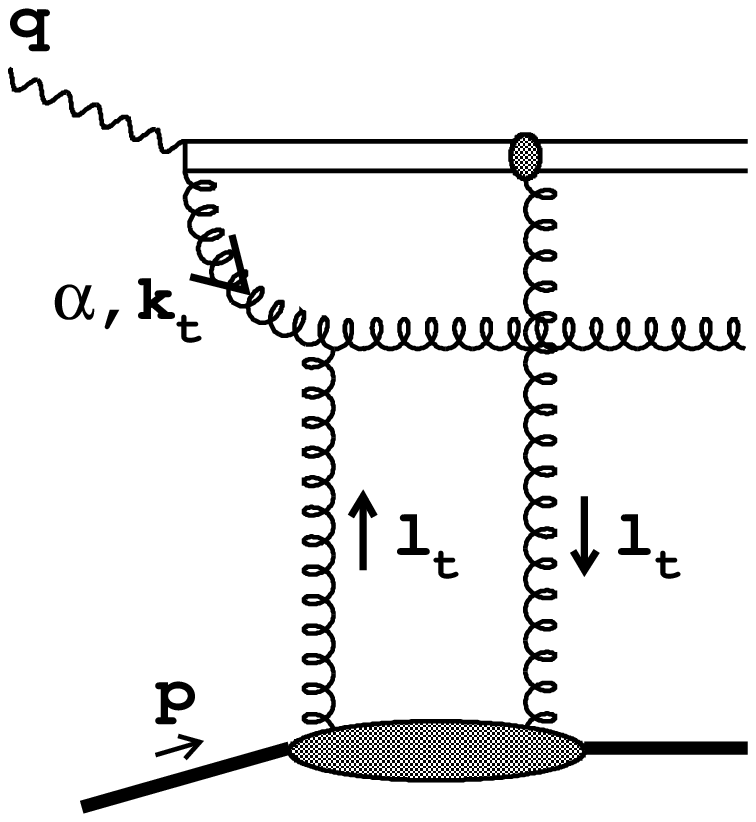,height=6cm}       
     \end{center}       
\vspace*{-1cm}       
\caption{\it Diffractive $q\bar{q}$ and $q\bar{q}g$ electroproduction mediated by the       
exchange of two gluons with transverse momenta $\pm {\bf \ell}_t$.  We       
view the photon as dissociating into either a quark dipole or an effective gluon dipole, made       
up of a gluon and a compact $q\bar{q}$ pair.        
\label{diag2}}       
\end{figure}       
      
The intuitive picture is as follows.  In the        
target (proton) restframe the photon dissociates into a $q\bar{q}$-pair       
far upstream the target. The $q\bar{q}$-pair       
may radiate a gluon, and the whole parton       
configuration scatters quasi-elastically off       
the proton via a two-gluon exchange. The timescale on which the fluctuation       
occurs is proportional to $1/(x\;m_p)$ where $m_p$ is the proton rest mass.       
At very small $x$ the fluctuation is long lived whereas the scattering       
is a sudden short impact of the $q\bar{q}$-pair or the        
$q\bar{q}g$-final state on the target. The impact changes the        
virtual into a real state but it does not change        
the position in impact parameter space which can be viewed as being frozen       
during the scattering.       
      
It is informative to show why the fluctuation and interaction timescales are so different.  For       
this it is convenient to use light-cone perturbation theory (see, for example, \cite{BL}) and to 
express the particle       
four momenta in the form      
\be      
\label{eq:b15}      
k_\mu \; = \; (k_+, k_-, \mbox{\boldmath $k$}_t)      
\ee      
where $k_\pm = k_0 \pm k_3$.  In this approach all the particles are on mass-shell      
\be      
\label{eq:c15}      
k^2 = k_+ k_- \: - \: k_t^2 \; = \; m^2,      
\ee      
and $k_+$ and $\mbox{\boldmath $k$}_t$ are conserved at each vertex.  In the proton rest       
frame we have      
\begin{eqnarray}      
\label{eq:d15}      
p_\mu \; = \; (m_p, m_p, \mbox{\boldmath $0$}), \quad\quad\quad & & q_\mu \; = \; (q_+, \:       
-Q^2/q_+, \: \mbox{\boldmath $0$}), \nonumber \\      
& & \\      
k_\mu \; = \; (k_+, \: m_T^2/k_+, \: \mbox{\boldmath $k$}_t), \quad\quad\quad & & \ell_\mu       
\; = \; (\ell_+, \:       
\ell_t^2/\ell_+, \: \mbox{\boldmath $\ell$}_t), \nonumber       
\end{eqnarray}      
with $m_T^2 = m^2 + k_t^2$ where $m$ is the mass of the quark.  According to the       
uncertainty principle the $\gamma^* \rightarrow q\bar{q}$ fluctuation time      
\be      
\label{eq:e15}      
\fl \tau_\gamma \: \sim \: \frac{1}{\Delta E} \; = \; \left | \frac{2}{q_- - k_{1-} - k_{2-}}       
\right |       
\; = \; \frac{2q_+}{Q^2 + m_T^2/\alpha + m_T^2/(1 - \alpha)} \: \simeq \: \frac{2q_+}{Q^2}       
\; = \; \frac{1}{m_p x}      
\ee      
at high $Q^2$, where $k_1$ and $k_2$ are the four momenta of the quark and antiquark.  The momentum fractions, 
shown in Fig.~3, are $\alpha = k_{1+}/q_+$ and $1 - \alpha = k_{2+}/q_+$.
The factor 2 in (\ref{eq:e15}) occurs because the energy $k_0 = (k_+ + k_-)/2$.  An       
estimate of the interaction time can be obtained from the typical time for the emission of a       
gluon of momentum $\ell$ from the quark $k_1$, say.  Then      
\be      
\label{eq:f15}      
\fl \tau_i \: \sim \: \left | \frac{2}{k_{1-} - k_{1-}^\prime - \ell_-} \right | \; = \; \left |       
\frac{2q_+}{m_T^2/\alpha - m_T^2/\alpha^\prime - \ell_t^2/(\alpha - \alpha^\prime)} \right |       
\: \simeq \: \frac{2 \ell_+}{\ell_t^2} \; = \; \frac{1}{m_p x_\ell}      
\ee      
since $\alpha - \alpha^\prime = \ell_+/q_+$ and $\alpha^\prime = k_{1+}^\prime/q_+$.  We can regard $x_\ell$ as the Bjorken variable       
for the gluon-proton interaction.  We are concerned with the kinematic region $x \ll x_\ell$       
where the leading $\log (x_\ell/x)$ approximation is appropriate, and so we have       
$\tau_\gamma \gg \tau_i$.      
      
One should note that the above picture is valid in        
a certain frame in combination with a certain physical gauge condition,       
$p\cdot A = 0$ ($p$ is the proton momentum and $A$ the gluon vector potential).      
In this gauge the parton shower evolves from the photon end while      
Bremsstrahlung from the proton is suppressed. One can turn       
to the Breit frame where the proton is        
fast moving and with it the Pomeron. The Pomeron in this frame is       
a long lived and very complicated virtual fluctuation. It is then       
more appropriate to use the gauge       
condition $q' \cdot A = 0$ ($q'=q+x p$ is the corresponding light-cone vector       
in the photon direction) under which the parton shower evolves from the Pomeron end.      
The results do not change      
but the physical interpretation is less appealing.       
      
The virtual fluctuation of the photon into a $q\bar{q}$-pair or      
a $q\bar{q}g$-final state is described by photon wave functions.      
The QCD-wave function for a $q\bar{q}$-pair       
has first been discussed in Refs.~\cite{NZ1,NZ2}. Similar results      
have already been derived within QED. In Ref.~\cite{BjSop}      
the fluctuation of a photon into a pair of muons has been studied      
as a model for deep inelastic scattering. The discussion of       
gluon radiation is a more recent development starting with       
Ref.~\cite{Rys}.       
      
The wave function for a $q\bar{q}$-pair depends on the polarization      
of the parent photon, whether it is transversely ($T$) or longitudinally ($L$) polarized.      
The key assumption in the whole approach is the eikonal-type      
coupling of the $t$ channel gluons with momentum $\ell$ to fast moving $s$ channel      
quarks and gluons such that
\begin{equation}
\langle p - \ell | \gamma^\mu | p \rangle \; \simeq \; 2p^\mu 
\label{eik}
\end{equation}
where $p$ and $p - \ell$ are the momenta of the fast incoming and outgoing partons. This approximation 
is correct at very high energies $(\ell \ll p)$.      
The derivation of the transverse and longitudinal wave functions, based      
on an explicit spinor representation, can be found in Ref.~\cite{Mue}      
or Ref.~\cite{IvWu}.  For a transversely polarized photon of, say, helicity $\lambda_\gamma       
= +1$ the wave functions $\Psi_{\lambda\lambda^\prime}$ are      
\begin{eqnarray}      
\label{eq:g15}      
\Psi_{+-}^T & = & \frac{\sqrt{2} e_q \alpha \mbox{\boldmath $k$}_t \cdot  
\mbox{\boldmath       
$\varepsilon$}       
(\lambda_\gamma = 1)}{k_t^2 + \alpha (1 - \alpha) Q^2}, \nonumber \\      
& & \\      
\Psi_{-+}^T & = & - \: \frac{\sqrt{2} e_q (1 - \alpha) \mbox{\boldmath $k$}_t \cdot       
\mbox{\boldmath       
$\varepsilon$} (\lambda_\gamma = 1)}{k_t^2 + \alpha (1 - \alpha) Q^2} \nonumber       
\end{eqnarray}      
where $\mbox{\boldmath $\varepsilon$}$ is the polarization vector of the virtual gluon,       
$\lambda, \lambda^\prime$ are the helicities of the quark, antiquark (with $\pm$ denoting       
helicity $\pm \frac{1}{2}$), and $e_q$ is the electric charge of the quark.  The general  
structure of       
(\ref{eq:g15}) is clear.  For a massless quark, helicity conservation at a quark-photon vertex       
requires $\lambda \neq       
\lambda^\prime$.  The momentum fractions $\alpha, (1 - \alpha)$ appear in the numerator       
because the photon helicity tends to follow the fast quark, antiquark.  Angular momentum       
conservation forbids collinear $q, \bar{q}$ production.  Hence the $k_t$ factor signifying a       
$P$-wave interaction.  Finally, the structure of the denominator has its origin in       
(\ref{eq:e15}) with $m_T^2 = k_t^2$.  The longitudinal wave functions (with       
$\lambda_\gamma = 0$) are found       
to be      
\be      
\label{eq:h15}      
\Psi_{+-}^L \; = \; \Psi_{-+}^L \; = \; \frac{2e_q \alpha (1 - \alpha) Q}{k_t^2 + \alpha (1 -       
\alpha) Q^2}.      
\ee      
      
The effective wave function for the $q\bar{q}g$-state requires some remarks.      
Since the photon does not couple to gluons directly it cannot decay      
into a gluon dipole. At large $Q^2$, however, the colour structure of the $q\bar{q}$ effectively       
combines into that of a gluon leading to a gluon dipole. This feature      
is based on the strong ordering of the transverse momenta       
in the leading log($Q^2$) approximation to diffractive DIS, that is, it is valid when $k_t \ll 
Q$ and $\alpha \ll 1$, and when the photons are transversely polarized. The contribution      
due to the quark-box can be factorized (as becomes more explicit later on) and we are left with the following      
expression for the effective gluon dipole wave function describing the $\gamma^* \rightarrow       
(q\bar{q}) g$ transition \cite{Wu,GBW}:      
\be      
\label{g-wave}      
\Psi_{q\bar{q}g} \; = \; \frac{1}{\sqrt{\alpha Q^2}}\;      
\frac{k_t^2\; (\mbox{\boldmath $\varepsilon$}_1 \cdot \mbox{\boldmath $\varepsilon$}_2)       
\: - \: 2 (\mbox{\boldmath $k$}_t \cdot \mbox{\boldmath $\varepsilon$}_1)       
(\mbox{\boldmath $k$}_t \cdot \mbox{\boldmath       
$\varepsilon$}_2)}{k_t^2 + \alpha Q^2},      
\ee      
where $\mbox{\boldmath $\varepsilon$}_1$ and $\mbox{\boldmath $\varepsilon$}_2$ are       
the polarization vectors of the $(q\bar{q})$ and $g$ forming the gluon dipole. Since $\alpha 
\ll 1$ we have replaced\footnote{In principle, there is a symmetry between $\alpha$ and $(1 -
\alpha)$, however, by convention, $\alpha$ is used to denote the small component.} $(1 - 
\alpha)$ by 1 in (\ref{g-wave}).
Otherwise the wave function is to be interpreted in analogy to eqs.~(\ref{eq:g15},  
\ref{eq:h15}).       
      
In order to ensure gauge invariance we have to consider all possible couplings      
of the two $t$ channel gluons to the quark- or \lq gluon\rq-dipole. In total there are      
four configurations of which only one is shown in Fig.~\ref{diag2}. This      
leads to the following expression      
\beqn\label{d-psi}      
D \Psi(\mbox{\boldmath $k$}_t,\mbox{\boldmath $\ell$}_t) & = & 2       
\;\Psi(\alpha,\mbox{\boldmath $k$}_t)\;-\;\Psi(\alpha,\mbox{\boldmath       
$k$}_t+\mbox{\boldmath $\ell$}_t)\;-\;      
\Psi(\alpha,\mbox{\boldmath $k$}_t-\mbox{\boldmath $\ell$}_t) \nonumber\\      
&\simeq&\cases{      
- \mbox{\boldmath $\ell$}_t^i \mbox{\boldmath $\ell$}_t^j\;\frac{\partial^2       
\Psi(\alpha,\mbox{\boldmath $k$}_t)}{\partial \mbox{\boldmath $k$}_t^i      
\partial \mbox{\boldmath $k$}_t^j}&\mbox{for $\ell_t \ll k_t$}, \cr      
2 \;\Psi(\alpha,\mbox{\boldmath $k$}_t)&\mbox{for $\ell_t \gg k_t$}.      
}      
\eeqn      
The amplitude for diffractive scattering is basically obtained      
by folding $D\Psi$ with the unintegrated gluon distribution $\cal{F}$:      
\beqn \label{amp}      
A^D &\sim& \int \frac{d^2\mbox{\boldmath $\ell$}_t}{\pi\;\ell_t^2}       
\;D\Psi(\mbox{\boldmath $k$}_t,\mbox{\boldmath $\ell$}_t)\;      
\alpha_S (\ell_t^2) {\cal F}(x_{\fP}, \ell_t^2)\;\;.      
\eeqn   
The form for $A^D$ is an example of the so-called \lq $k_t$-factorization theorem\rq~\cite{Catani}, here       
applied to the transverse momentum $\ell_t$.  The distribution ${\cal F}$, unintegrated over       
$\ell_t^2$,contains all the details of the coupling of the two $t$ channel gluons to the proton       
as indicated by the bubble in Fig.~\ref{diag2}\footnote{To be precise we should use the \lq   
skewed\rq~gluon distribution since the momentum flowing along the exchanged gluon lines   
in Fig.~3 are not exactly equal and opposite, see Section 11.}.  This gluon distribution ${\cal  
F} (x, \ell_t^2)$       
is a universal function applicable to all hard scattering processes involving the proton and,       
indeed, the gluon is the dominant parton distribution at small $x$.  Integrating ${\cal F} (x,       
\ell_t^2)$ over $\ell_t^2$ gives the conventional gluon distribution $xg (x)$.      
      
We have now, in principle, all the \lq\lq QCD-based\rq\rq~ingredients to write down the       
diffractive cross section. Before we do that we comment on      
the role of the wave function for the $\beta$-spectrum. It has been      
demonstrated in Ref.~\cite{BEKW} that the basic shape of the $\beta$-spectrum      
arises from the wave functions. It depends only weakly on the details      
of the unintegrated distribution $\cal{F}$. Looking back at      
(\ref{d-psi}) we see a \lq hard' ($\ell_t \ll k_t$) and a \lq soft' ($\ell_t \gg k_t$)\footnote{In the   
\lq soft' limit the two gluons couple to the same  
quark line, that is they act like a single pseudo exchange particle with $C=1$  
and therefore give the same result as in (\ref{DL-eq}) \cite{LN}.} limit for       
$D\Psi$.  The typical scale for $\ell_t$ is given by the dynamics inside the proton.       
The hard limit is represented by the second derivative of the wave functions      
and the soft limit by the wave function itself.  The second derivatives are       
\beqn      
\delta^{i j}\;\frac{\partial^2 \Psi^T_{+-}}{\partial \mbox{\boldmath $k$}_t^i      
\partial \mbox{\boldmath $k$}_t^j}&=&      
\alpha(1-\alpha)Q^2 \;\frac{\sqrt{2}\;e_q \;(1 - \alpha) \;\mbox{\boldmath $k$}_t \cdot       
\mbox{\boldmath $\varepsilon$}       
(\lambda_\gamma = 1)}{[k_t^2+\alpha (1-\alpha)Q^2]^3},  \label{eq:26}\\      
\delta^{i j}\;\frac{\partial^2 \Psi^L_{+-}}{\partial \mbox{\boldmath $k$}_t^i      
\partial \mbox{\boldmath $k$}_t^j}&=&      
4\;e_q\;\alpha(1-\alpha)\;Q \;\frac{k_t^2-\alpha(1-\alpha)Q^2}      
{[k_t^2+\alpha (1-\alpha)Q^2]^3}, \label{l-second}\\      
\delta^{i j}\;\frac{\partial^2 \Psi_{q\bar{q}g}}{\partial \mbox{\boldmath $k$}_t^i      
\partial \mbox{\boldmath $k$}_t^j}&=&      
\frac{k_t^2 (\mbox{\boldmath $\varepsilon$}_1 \cdot \mbox{\boldmath  
$\varepsilon$}_2) \: - \: 2 (\mbox{\boldmath $k$}_t \cdot \mbox{\boldmath 
$\varepsilon$}_1) (\mbox{\boldmath $k$}_t \cdot \mbox{\boldmath       
$\varepsilon$}_2)}{\sqrt{\alpha Q^2}}\;\frac{k_t^2+3\alpha Q^2}      
{[k_t^2+\alpha Q^2]^3}. \label{g-second}      
\eeqn      
Moreover, since the mass $M$ is formed from two subsystems with $k_+$ components       
$\alpha q_+$ and $(1 - \alpha) q_+$ we have      
\be      
\label{eq:z}      
M^2 \; = \; (k_1 + k_2)^2 \; = \; (q_+, \: k_t^2/(\alpha (1-\alpha) q_+), \: \mbox{\boldmath       
$0$})^2      
\ee      
and hence      
\be\label{kin1}      
k_t^2\;=\;M^2\;\alpha(1-\alpha).      
\ee      
Therefore for both the \lq hard' and \lq soft' regimes, in the limit of small      
diffractive masses, $M \ll Q$, and keeping $\alpha$ fixed, we have for      
\beqn      
\Psi_{q\bar{q}}^T\hspace*{0.4cm},\hspace*{0.5cm} \partial^2 \Psi^T_{q\bar{q}}       
\hspace{0.5cm}&\sim& \hspace{0.5cm}k_t \;\sim \;M, \label{eq:x} \\      
\Psi_{q\bar{q}}^L \hspace*{0.4cm},\hspace*{0.5cm} \partial^2 \Psi^L_{q\bar{q}}       
&\sim&\hspace{0.5cm} {\rm constant}, \label{eq:xx} \\      
\Psi_{q\bar{q}g} \hspace*{0.2cm},\hspace*{0.5cm} \partial^2 \Psi_{q\bar{q}g} &\sim&       
\hspace{0.5cm}k^2_t \;\sim \;M^2. \label{eq:xxx}      
\eeqn      
One also finds from the analysis of the wave function that the longitudinal      
part is a higher twist contribution, i.e. the longitudinal structure function      
is suppressed by an extra power in $Q^2$ at fixed $\beta$, where
\be
\label{eq:a37}
\beta \; \equiv \; \frac{x}{x_\funp} \; = \; \frac{Q^2}{Q^2 + M^2}.
\ee
One can show this      
result by substituting $\alpha$ in (\ref{l-second}) and (\ref{eq:26})      
by $k_t^2/M^2$ according to (\ref{kin1}) assuming that $\alpha$ is small.      
One then obtains for (\ref{l-second}) an extra factor $k_t/Q$ as compared to      
(\ref{eq:26}).        
      
The photon wave function determines the general structure of the $\beta$ spectrum.  We have 
already noted that at large masses $M$ the       
contribution to the inclusive diffractive process $\gamma^* p \rightarrow Xp$ arising from       
$q\bar{q}g$ production dominates over $q\bar{q}$ even though it is higher order in       
$\alpha_S$, since it contains gluon exchange as opposed to quark exchange (see Fig.~3).  In       
summary, (\ref{eq:x})--(\ref{eq:xxx}) indicate that the $\beta$-distribution      
has a distinct separation into three regions of small, medium and large      
$\beta$ (large, medium and small $M$) where each of the three contributions,      
$q\bar{q}g$, transverse and longitudinal $q\bar{q}$ respectively dominates. The fact      
that the longitudinal part is non-negligible also indicates the importance      
of higher twist terms even at fairly large $Q^2$.       
One of the important conclusions in Ref.~\cite{BEKW} was the observation that      
the wave function for $q\bar{q}g$ results in a rather soft gluon distribution (strongly 
decreasing as $\beta \rightarrow 1$), see (\ref{eq:xxx}).      
It has been shown that a parameterization based on the wave function formalism      
provides a good description of the data without the need of a very hard,      
singular gluon distribution (strongly peaked as $\beta \rightarrow 1$) as suggested by H1~\cite{H1diff}.       
      
We use the $\ell_t$ factorization formula (\ref{amp}) to obtain the explicit form of the  
diffractive       
structure functions.  First we note that the gluon distribution ${\cal F}$ is independent of the       
azimuthal angle of $\mbox{\boldmath $\ell$}_t$, as $t \rightarrow 0$.  We can therefore 
easily perform the       
integration over this angle.  We then take the square of the amplitude and integrate over $t$       
assuming a simple exponential form, $\exp (-B_D | t | )$, where the diffractive slope $B_D$       
is known from experiment.  The final result for the diffractive structure function $F_2^{D       
(3)} (\beta, Q^2, x_\funp)$ is the sum of the three contributions \cite{GBW}
\beqn      
\fl & & x_{\fP}F_{T,q\bar{q}}^D(\beta, Q^2, x_{\fP})  =  \frac{1}{96 B_D}\; \sum_q e^2_q      
\;\frac{Q^2}{1-\beta}\;\int_0^1d\alpha\;      
[\alpha^2+(1-\alpha)^2] \; \label{fqqt} \nonumber \\       
\fl & & \quad\quad\quad \times       
\left\{\int\frac{d\ell_t^2}{\ell_t^2} \alpha_S {\cal F}(x_{\fP},\ell_t^2)      
\left[1-2\beta\;+\;      
\frac{\ell_t^2 - (1-2\beta)\;k^2}{\sqrt{(\ell_t^2 + k^2)^2      
-4 (1-\beta)\;\ell_t^2 \;k^2}} \right]\right\}^2, \\      
\fl & &  x_{\fP}F_{L,q\bar{q}}^D(\beta, Q^2, x_{\fP}) \label{fqql}      
 =  \frac{1}{6 B_D} \; \sum_q e^2_q \;      
\int_0^1d\alpha\;k^2\;\beta^2\;\; \nonumber \\       
\fl & & \quad\quad\quad \times       
\left\{\int\frac{d\ell_t^2}{\ell_t^2} \alpha_S{\cal F}(x_{\fP},\ell_t^2)      
\left[1\;-\;\frac{k^2}      
{\sqrt{(\ell_t^2+k^2)^2 -4 (1-\beta)\;\ell_t^2\;k^2}} \right]\right\}^2,      
\\      
\fl & & x_{\fP}F_{q\bar{q}g}^D(\beta, Q^2, x_{\fP}) \label{fqqg}      
 =  \frac{9 \;\beta}{64 B_D}\, \sum_q e_q^2 \;\int_0^{Q^2} dk^2\;      
\frac{\alpha_S}{2\pi}\;\ln\left(\frac{Q^2}{k^2}\right)\nonumber \\       
\fl & & \quad\quad\quad \times       
\;\int_\beta^1 \frac{dz}{z^2\,(1-z)^2}\;       
\left[\left(1-\frac{\beta}{z}\right)^2      
+\left(\frac{\beta}{z}\right)^2\right]      
\left \{ \int\frac{d\ell_t^2}{\ell_t^2}\;\alpha_S{\cal F}(x_{\fP},\ell_t^2) \right .     
\;\;  \\      
\fl & & \quad\quad\quad \times  \left[z^2+(1-z)^2+\frac{\ell_t^2}{k^2}-      
\left . \frac{[(1-2z)k^2-\ell_t^2]^2+2z(1-z)k^4}      
{k^2\sqrt{(\ell_t^2+k^2)^2-4(1-z)\;\ell_t^2\;k^2}}      
\right]\right \}^2 \;\;.\nonumber      
\eeqn      
We have changed variables in the square brackets       
using ( 
\ref{kin1}), and in addition we have introduced      
\beqn\label{kin2}      
k^2\;=\;\frac{k_t^2}{1-\beta}\hspace*{1cm}\mbox{for $q\bar{q}$}, \\      
k^2\;=\;\frac{k_t^2}{1-z}\hspace*{1cm}\mbox{for $q\bar{q}g$}\;\;.\nonumber      
\eeqn      
Note that the infrared cut-off is hidden in the unintegrated gluon ${\cal F}$ of (\ref{amp}).  It 
is specified by the size of the proton.  Eqs.~(\ref{fqqt}--\ref{fqqg}) are therefore safe in the 
infrared limit, $k^2 \rightarrow 0$.  The variable $z$ appears for $q\bar{q}g$ instead of 
$\beta$ because we still      
have to convolute with the quark box. The convolution is apparent in terms of the       
Altarelli-Parisi splitting function, describing the $g\rightarrow q\bar{q}$ transition (that is the 
second factor in the second line of (\ref{fqqg})).      
The $z$ describes the relative momentum fraction of the gluon with respect      
to the Pomeron.      
      
The unintegrated gluon distribution $\cal{F}$ is the only quantity in       
(\ref{fqqt}-\ref{fqqg}) still to be determined in order to perform       
a numerical analysis. It can      
be obtained from standard parameterizations for parton distributions or      
by using a particular model as in Ref.~\cite{GBW}.      
 \begin{figure}[htb]        
  \begin{picture}(15,6)(0,0.5)      
\put(0.7,6){$x_{\fP} F_2^D$}      
\put(2.6,2.7){$q\bar{q}g$}      
\put(7.75,3){$q\bar{q}g$}      
\put(4.4,2.5){T}      
\put(9.3,2.6){T}      
\put(5.5,1.75){L}      
\put(10.6,1.5){L}      
\put(1,0){\epsfig{figure=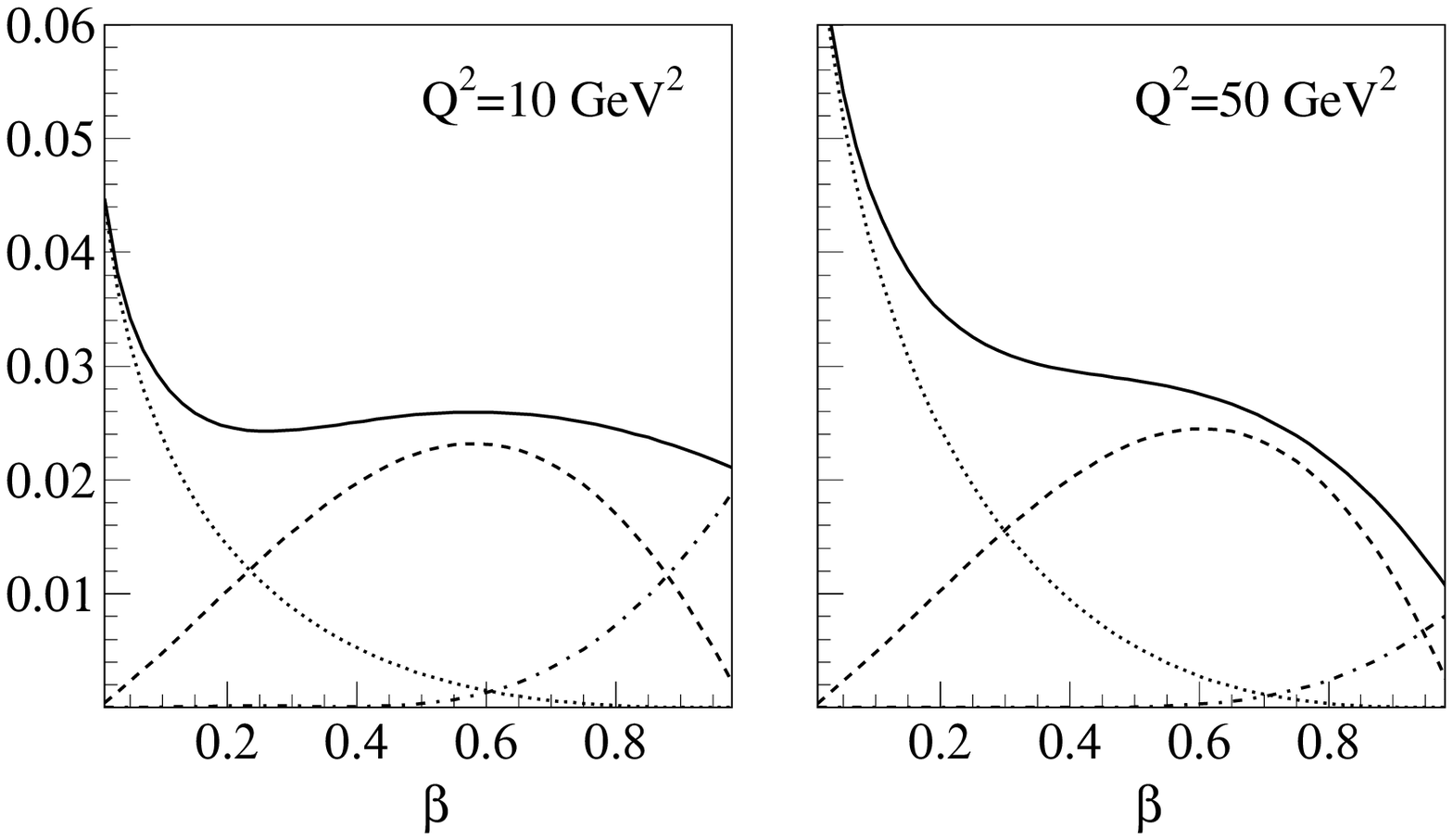,height=7cm}\hspace*{-1cm}}       
\end{picture}      
\caption{\it The diffractive structure function $F_2^D$ and its three components      
plotted versus $\beta$ for a fixed $x_{\fP}=0.005$.      
Dashed line: transverse $q\bar{q}$, dashed-dotted line: longitudinal $q\bar{q}$      
and dotted line:  $q\bar{q}g$.  \label{ddrev_plot}}       
\end{figure}        
Fig.~\ref{ddrev_plot} shows how the three contributions, $q\bar{q}g$, $q\bar{q}$      
from transversely and longitudinally polarized photons, occupy the three different       
regimes at low, medium and high      
$\beta$. This decomposition is in large due to the property of the wave fuctions      
as argued previously. The precise form of the parameterization for       
$\cal{F}$ has only little influence on the $\beta$ spectrum. The relative normalization is       
given by colour      
factors. Since the colour factor for $q\bar{q}g$ is much bigger than for $q\bar{q}$, it is       
another reason why it becomes important despite the fact that it is a higher order contribution      
($\alpha_S$ was set to 0.2). One notices the drop of the curve for $F_2^D$ at large $\beta$      
when $Q^2$ is increased. This effect is due to the higher twist nature of the      
longitudinal contribution as discussed earlier. The rise at small $\beta$, on the other hand,      
is driven by the logarithm in $Q^2$ which results from the phase space integration of the       
quark box.      
      
The $x_{\fP}$-distribution depends very much on the parameterization that one chooses for   
the gluon distribution $\cal{F}$. In general, since we are considering a perturbative approach,   
the slope is much steeper as compared to a soft approach. As an example for a soft approach  
one can take the model of Ref.~\cite{LN} which is based on nonperturbative gluon exchange.  
It leads to a much shallower $x_{\fP}$-distribution \cite{Diehl,Diehl2}.

\section{Impact parameter representation}      
It is informative to recast the formulae derived so far in impact parameter space.      
In this way we can see the physical      
picture described earlier in which the position of the partons       
in impact parameter space is frozen during the scattering. This property,       
and the fact that the colour-dipole approach to small $x$ physics      
in Ref.~\cite{MuePat,NNCD,NZ3,Pesch} lives in impact parameter space, has       
led to its increased popularity. Moreover, the concept of the      
dipole cross section can be generalized from two-gluon exchange to multi-gluon      
exchange.       
The disadvantage, however, is the need to transform back to momentum space when       
exclusive distributions such as the $\beta$-spectrum      
are studied.      
      
We transform the wave functions $\Psi (\mbox{\boldmath $k$}_t, \alpha)$ of     
(\ref{eq:g15})--(\ref{g-wave}) to       
impact parameter space using the Fourier transformation       
\beqn      
\psi (\mbox{\boldmath $r$},\alpha)&=&\int \frac{d^2 \mbox{\boldmath $k$}_t}{(2\pi)^2}       
\;e^{i\;\mbox{\boldmath $k$}_t \cdot       
\mbox{\boldmath $r$}}\; \Psi(\mbox{\boldmath $k$}_t,\alpha)  
\label{ftwf}    
\eeqn      
where $r$ denotes the separation of the dipole in impact parameter space.  Carrying out the       
integration we find      
\beqn\label{impact-t-wave}      
\fl \psi_{+-}^T(\mbox{\boldmath $r$},\alpha)&=& \frac{\sqrt{2}i e_q}{2 \pi}\;      
\alpha^{\frac{3}{2}}(1-\alpha)^{\frac{1}{2}}\; Q\;K_1(\sqrt{\alpha(1-\alpha) Q^2 r^2})\;      
\hat{\mbox{\boldmath $r$}} \cdot \mbox{\boldmath $\varepsilon$} (\lambda_\gamma = 1),       
\\      
\fl \psi_{-+}^T (\mbox{\boldmath $r$}, \alpha) & = & - \: \frac{\sqrt{2}ie_q}{2 \pi} \:     
\alpha^{\frac{1}{2}} (1 - \alpha)^{\frac{3}{2}} \: Q \: K_1 (\sqrt{\alpha (1 - \alpha) Q^2     
r^2}) \: \hat{\mbox{\boldmath $r$}} \cdot \mbox{\boldmath $\varepsilon$}  
(\lambda_\gamma = 1),  \nonumber \\    
\fl \psi_{+-}^L (\mbox{\boldmath $r$},\alpha)&=& \psi_{-+}^L (\mbox{\boldmath $r$},     
\alpha) \; = \; \frac{e_q}{\pi}\;      
\alpha(1-\alpha)\; Q\;K_0(\sqrt{\alpha(1-\alpha) Q^2 r^2}), \label{impact-l-wave} \\      
\label{impact-g-wave}      
\fl \psi_{q\bar{q}g}(\mbox{\boldmath $r$}, \alpha)      
&=&-\frac{1}{2\pi}\left(\mbox{\boldmath $\varepsilon$}_1 \cdot \mbox{\boldmath       
$\varepsilon$}_2 \: - \: 2 \hat{\mbox{\boldmath $r$}} \cdot \mbox{\boldmath       
$\varepsilon$}_1 \: \hat{\mbox{\boldmath $r$}} \cdot \mbox{\boldmath $\varepsilon$}_2       
\right)\;      
\sqrt{\alpha Q^2}\;K_2(\sqrt{\alpha Q^2 r^2})\;\;,      
\eeqn      
where the $K_n$ are modified Bessel functions of the second kind.  We can use these wave       
functions to evaluate the diffractive structure functions.  For example we find      
\beqn \label{impact-fqqt}      
\fl & & x_{\fP}F_{T, q\bar{q}}^D  =  \frac{3}{64 \pi^5 B_D}\;\sum_q e^2_q       
\;\frac{\beta^2}{(1-\beta)^3}\; \int \frac{d^2\mbox{\boldmath $k$}_t}{(2\pi)^2}\;\;k_t^4 \:       
\frac{1 - 2a^2 k_t^2/Q^2}{\sqrt{1-4a^2 k_t^2/Q^2}} \: \Theta (1 - 4a^2 k_t^2/Q^2)       
\nonumber \\      
\fl & & \quad\quad \times \int d^2\mbox{\boldmath $r$}\;\int d^2\mbox{\boldmath       
$r$}'\;\;e^{i\;\mbox{\boldmath $k$}_t \cdot (\mbox{\boldmath       
$r$}-\mbox{\boldmath $r$}^\prime)} \;\;\hat{\sigma}(r,x_{\fP})\;      
\hat{\sigma}(r',x_{\fP})\: \hat{\mbox{\boldmath $r$}} \cdot \hat{\mbox{\boldmath       
$r$}}^\prime \: K_1 (a k_t r) \: K_1 (a k_t r^\prime),       
\eeqn      
where $a^2 \equiv \beta/(1-\beta)$ and where we have introduced the dipole cross section      
\beqn\label{sig_f}      
\hat{\sigma}(r,x)&=&\frac{4 \pi}{3}\;      
\int \frac{d^2 \mbox{\boldmath $\ell$}_t}{\ell_t^2} \;      
\left[ \;1 - e^{i\;\mbox{\boldmath $r$} \cdot \mbox{\boldmath $\ell$}_t}\; \right]\;       
\alpha_S\,{\cal F}(x,\ell_t^2) \nonumber \\       
&=&\frac{4 \pi^2}{3}\; \int \frac{d \ell^2_t}{\ell_t^2} \;      
\left[\;1-J_0(\ell_t r)\;\right]\;\alpha_S \,{\cal F}(x,\ell_t^2)\;\;.      
\eeqn        
Inserting (\ref{sig_f}) into (\ref{impact-fqqt}) and computing the integral over       
$\mbox{\boldmath $r$}$ or $\mbox{\boldmath $r$}^\prime$ one directly gets back to the       
expression in (\ref{fqqt}).  $\hat{\sigma} (r,x)$ is the effective       
inclusive cross section for the scattering of a $q\bar{q}$ system, with $q, \bar{q}$ transverse       
separation $r$, on the proton at energy $\sqrt{s}$, where $s = M^2/x$.  The diffractive or       
elastic scattering amplitude is given in terms of $\hat{\sigma}$, via the optical theorem.        
Hence the $\hat{\sigma}\hat{\sigma}$ structure of the observable (\ref{impact-fqqt}).      
      
One important motivation for using the impact parameter representation      
is the factorization into the dipole wave function      
and dipole cross section, which is equivalent to $k_t$-factorization      
in momentum space. The impact parameter does not change in the course       
of the scattering, i.e. the incoming and outgoing states have the same      
impact parameter. This feature is clearly visible in the $\beta$- or       
$M$-integrated version of (\ref{impact-fqqt}).      
When $\beta$ is changed back to $\alpha$ one obtains      
\beqn\label{impact-fqqt-int}      
\fl & & F_{T, q\bar{q}}^{D (2)}(x, Q^2)  = \frac{3 Q^2}{128 \pi^5 B_D}\;\sum_q e_q^2\;      
\int_0^1d\alpha\;[\alpha^2+(1-\alpha)^2]\;\; \nonumber \\      
\fl & & \quad\quad\quad\quad\quad\quad \times \alpha(1-\alpha)\; Q^2\;\int  
d^2\mbox{\boldmath $r$}\;      
K_1^2(\sqrt{\alpha(1-\alpha) Q^2 r^2})\; \hat{\sigma}^2(r,x)      
\eeqn      
which involves the square of the dipole cross section $\hat{\sigma}$.  That is $r = r^\prime$,       
and the impact parameter is unchanged by the interaction with the proton.  Mass eigenstates, however, 
are not eigenstates of the impact parameter (see (\ref{impact-fqqt})) \cite{Peschanski}.
       
\section{Factorization in $\gamma^* p \rightarrow Xp$ and diffractive parton distributions       
\label{fact}}      
A general proof of collinear factorization in diffractive deep      
inelastic scattering has been given in Ref.~\cite{Col} (see also      
\cite{BerSop}).      
In Ref.~\cite{Col} it has also been stated that factorization is      
violated in diffractive hadron-hadron scattering. We will demonstrate this in section  
\ref{sect-hadron}.  In general we cannot factorize diffractive cross       
sections into a convolution of a \lq hard\rq~partonic subprocess with universal diffractive       
parton distributions.  On the other hand the factorization proof allows us to introduce such       
distributions for diffractive deep inelastic scattering processes.      
      
We can demonstrate from the expressions for the diffractive structure functions, that we       
can extract diffractive parton distributions which are in      
agreement with collinear factorization. For $q\bar{q}$ production we only have to strip      
off terms from $F_{T,q\bar{q}}^D$ which are of higher twist nature (such as $a^2 k_t^2/Q^2$ in 
(\ref{impact-fqqt})). This of course means  
that the longitudinal contribution is obsolete in this context.  In direct analogy to the parton  
model of ordinary DIS, (\ref{eq:b3}), for diffractive DIS processes we write      
\be      
\label{eq:41a}      
F_{T, q\bar{q}}^{D (3)} \; = \; \sum_q \: e_q^2 \: \beta \: q^D (\beta)      
\ee      
where we have introduced the diffractive quark distribution.  Notice that $\beta$ plays the       
role of the Bjorken $x$ variable for diffractive processes.  Identification (\ref{eq:41a}) is     
made for fixed values of $x_\funp$ and $Q^2$.  It is often said that $q^D (\beta)$ is       
the quark distribution of the \lq hard\rq~Pomeron with the quark carrying a fraction $\beta =       
x/x_\funp$ of its momentum.  Of course to introduce such a picture we would have to specify     
the probability of finding the Pomeron in the proton, and to assume that the Pomeron is a real     
particle (that is a pole in the $t$ plane or complex angular momentum plane).  The ambiguities 
and difficulties of interpreting $q^D (\beta)$ as the quark     
distribution of the Pomeron are discussed in Section 14.  By comparing (\ref{eq:41a})       
with the leading twist part of (\ref{impact-fqqt}) we may introduce the diffractive quark distribution (see also \cite{BH})
\beqn\label{quark-distr}      
\fl q^D(\beta)&=&\frac{3}{128 \pi^5 x_\funp B_D}\;\frac{\beta}{(1-\beta)^3}\;      
\int \frac{d^2\mbox{\boldmath $k$}_t}{(2\pi)^2}\;\;k_t^4  \: \int  
d^2\mbox{\boldmath $r$}\;\int d^2\mbox{\boldmath $r$}'\;\;e^{i\;\mbox{\boldmath $k$}_t  
\cdot (\mbox{\boldmath $r$}-\mbox{\boldmath $r$}^\prime)} \nonumber \\      
\fl & \times& \hat{\sigma}(r,x_{\fP})\; \hat{\sigma}(r',x_{\fP}) \: \hat{\mbox{\boldmath  
$r$}}  
\cdot \hat{\mbox{\boldmath $r$}}^\prime \: K_1 (a k_t r) \: K_1 (a k_t r^\prime),
\eeqn      
where recall $a^2 \equiv \beta/(1 - \beta)$.  Similarly using the $q\bar{q}g$ structure function 
$F_{q\bar{q}g}^D$ we may introduce the diffractive gluon distribution\footnote{The Operator Product 
Expansion allows diffractive parton distributions to be introduced consistently to any order in 
$\log  Q^2$.  According to \cite{HAUT} the initial distributions of (\ref{quark-distr}) and (\ref{gluon-distr}) 
are valid beyond leading $\log Q^2$.}
\beqn\label{gluon-distr}      
\fl  g^D(\beta) &=& \frac{81}{256 \pi^5 x_\funp B_D}\;\frac{\beta}{(1-      
\beta)^3}\; \int\frac{d^2 \mbox{\boldmath $k$}_t}{(2\pi)^2}\;k_t^4 \;\int       
d^2\mbox{\boldmath $r$}\;\int       
d^2\mbox{\boldmath $r$}'\;\;e^{i\;\mbox{\boldmath $k$}_t \cdot (\mbox{\boldmath $r$}-      
\mbox{\boldmath $r$}^\prime)} \label{gluon} \nonumber \\      
\fl &\times& \hat{\sigma}(r,x_{\fP})\;\hat{\sigma}(r',x_{\fP})\;       
(\delta^{mn} - 2       
\hat{r}^m \hat{r}^n) \: (\delta^{mn} - 2 \hat{r}^{\prime m}       
\hat{r}^{\prime n}) \: K_2 (ak_t r) \: K_2 (ak_t r^\prime)      
\eeqn      
with $m, n = 1,2$.  There are differences between using the full expression       
(\ref{impact-fqqt}) for $x_\funp       
F_T^D$ and using $q^D$ of (\ref{quark-distr}) in (\ref{eq:41a}).  One difference is in the       
form of the $k_t$       
integration; in particular in the absence of an upper limit on $k_t$ in (\ref{quark-distr}).  This       
means that energy-momentum      
conservation is obviously violated. The integral is nonetheless       
well defined. Imposing the true kinematical cut-off would \lq only' lead to       
higher twist corrections which in the context of collinear factorization      
are subleading.      
The importance of the longitudinal contribution in  comparison with the data,      
however,      
already indicates that higher twist corrections are not completely negligible.      
Another issue is the factorization scale $\mu^2$.  A dependence of the diffractive parton 
distributions on the scale $\mu^2$ is introduced by evolution, where (\ref{quark-distr}) and 
(\ref{gluon-distr}) serve as the initial distributions.  The initial scale $Q_0^2$, however, 
cannot exactly be determined, which introduces a considerable uncertainty for the prediction.

In the conventional parton picture there is always a soft remnant which carries the opposite 
colour to that of the elastically scattered parton.  For the dominant diffraction contribution the 
picture is the same.  There are, however, events with no soft remnant.  One example is the 
exclusive $q\bar{q}$ final state with large $k_t$.  Although subleading, these events constitutes 
an example which breaks factorization.  An important signal for these events is the azimuthal angle 
distribution of the $q$ and $\bar{q}$ jets \cite{BELW,Diehl}.
      
This discussion is not intended to disprove factorization but to point out       
some of the limitations. Diffraction has the unique property that the whole      
event is contained in the detector, i.e. the \lq soft' remnant is included.      
Factorization focuses on the hard subprocess and basically disregards the soft      
remnant. The approach we are pursuing based on Feynman diagrams       
can overcome part of these caveats.  It offers a complete description      
of the final state where the remnant is represented by a quark or gluon.      
      
\section{Triple Regge limit \label{triple-regge}}      
The $q\bar{q}g$ contribution in the triple Regge limit, i.e.      
$M^2\gg Q^2$, can be calculated assuming the strong      
ordering of the longitudinal momentum components. We can relax the ordering      
condition on the transverse momenta instead.  A further requirement      
in this approach is the strong hierarchy between the      
large diffractive mass $M$ and the transverse momenta $k_t$ of the      
quarks or the gluon, i.e. $M\gg k_t$.       
 
The diffractive structure function in this limit reads \cite{BJW}:      
\be  \label{fqqgtr}      
\fl x_{\fP}F_{q\bar{q}g}^D \; = \;  \sum_q e_q^2\; \frac{9\; \alpha_S \;       
Q^2}{512\;\pi^3\;B_D}\;\int_0^1 d \alpha \left[\alpha^2 + (1-\alpha)^2\right] \int       
d^2\mbox{\boldmath $k$}_1 d^2\mbox{\boldmath $k$}_2 \;M_{ij} M_{ij}^*,    
\ee      
where $i, j = 1, 2$ are vector components in the transverse plane.  The amplitude is given by      
\begin{eqnarray} \label{mil}      
\fl & & M_{ij} =  \int \frac{d^2\mbox{\boldmath $\ell$}}{\pi \ell^2}\;       
\alpha_S{\cal F}(x_{\fP}, \ell^2) \; \biggl \{ \left( \frac{\mbox{\boldmath       
$\ell$} + \mbox{\boldmath $k$}_1 + \mbox{\boldmath $k$}_2}{D(\mbox{\boldmath $\ell$}       
+ \mbox{\boldmath $k$}_1 +       
\mbox{\boldmath $k$}_2)} + \frac{\mbox{\boldmath $k$}_1 + \mbox{\boldmath       
$k$}_2}{D(\mbox{\boldmath $k$}_1 + \mbox{\boldmath $k$}_2)}\right. \nonumber \\      
\fl && \quad\quad\quad\quad \left.  - \frac{\mbox{\boldmath $k$}_1- \mbox{\boldmath       
$\ell$}}{D(\mbox{\boldmath $k$}_1-\mbox{\boldmath $\ell$})} -      
\frac{\mbox{\boldmath $k$}_1}{D(\mbox{\boldmath $k$}_1)} \right)_i       
\left ( \frac{\mbox{\boldmath $\ell$ 
} + \mbox{\boldmath $k$}_2}{(\mbox{\boldmath $\ell$}       
+ \mbox{\boldmath $k$}_2)^2} -       
\frac{\mbox{\boldmath $k$}_2}{\mbox{\boldmath $k$}_2^2} \right)_j \: + \:       
\left ( \mbox{\boldmath $\ell$} \to -\mbox{\boldmath $\ell$} \right) \biggr \},      
\eeqn      
where we have introduced the abbreviation      
\be      
\label{eq:47}      
D(\mbox{\boldmath $k$}) \; = \; \alpha(1-\alpha)Q^2 + \mbox{\boldmath $k$}^2,      
\ee      
\begin{figure}[htb]       
  \vspace*{-.5cm}      
     \begin{center}      
         \epsfig{figure=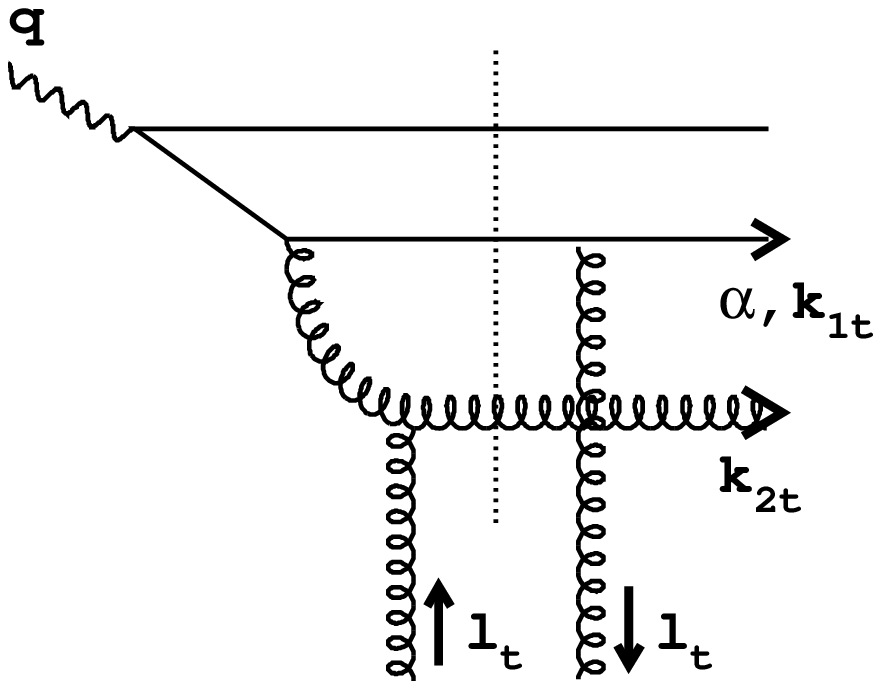,height=5cm}\hspace*{-1cm}      
         \epsfig{figure=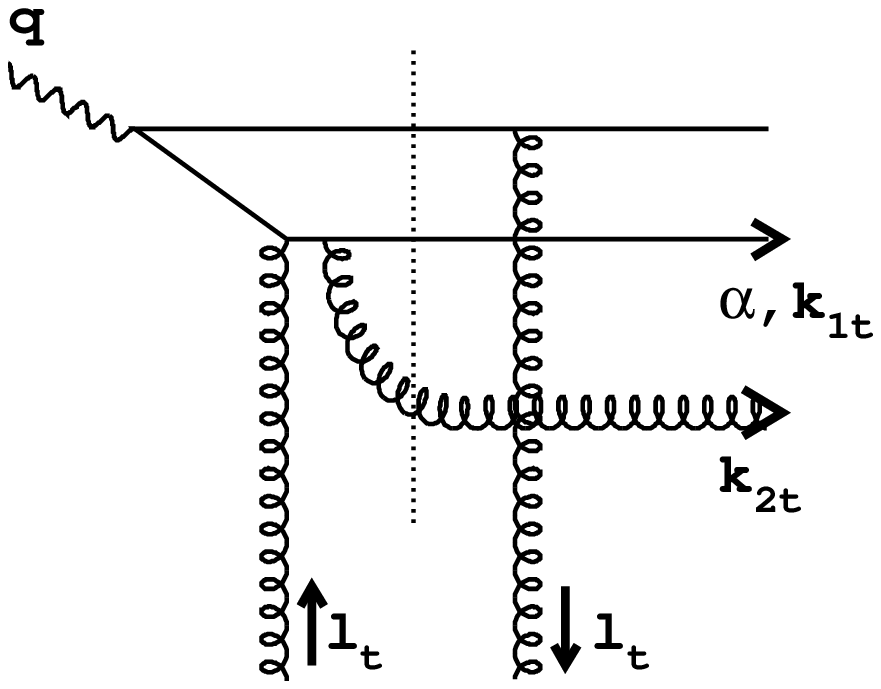,height=5cm}      
     \end{center}      
\vspace*{-1cm}      
\caption{\it Two typical diagrams contributing to diffractive $q\bar{q}g$ production.  In the     
first diagram the momentum of the off-shell quark is $\mbox{\boldmath $k$}_1 +     
\mbox{\boldmath $k$}_2$, and in the second is $\mbox{\boldmath $k$}_1 +     
\mbox{\boldmath $k$}_2 + \mbox{\boldmath $\ell$}$. The dotted line denotes a cut through the diagram 
(giving twice the imaginary part of the amplitude, which dominates over the real part).  \label{diag6}}      
\end{figure}      
and where $\mbox{\boldmath $k$}_1$ and $\mbox{\boldmath $k$}_2$ are the transverse       
momenta of the quark and the gluon respectively (see Fig. \ref{diag6}).  The first factor     
$(\cdots)_i$ in (\ref{mil}) represents the different forms of the propagator of the off-shell     
quark in diagrams such as those in Fig.~5.  The second factor $(\cdots)_j$ is associated with     
the coupling of the $s$ channel gluon (to be precise it is the BFKL vertex in the $p \cdot A =     
0$ gauge, which also encompasses gluon Bremsstrahlung emission); it also includes the   
gluon propagator.  Unlike (\ref{g-wave}), here $\alpha$ is the longitudinal momentum   
fraction of the quark and not the       
gluon.   The longitudinal contribution to $F_{q\bar{q}g}^D$ has been neglected here (see       
Ref.~\cite{BJW}).  One can also study the dependences on the azimuthal angle of       
$\mbox{\boldmath $k$}_1$ and $\mbox{\boldmath $k$}_2$.     
 
For practical purposes the representation in momentum space is inevitable. The previous     
result becomes more compact, though, if one transforms (\ref{mil}) into impact parameter     
space:      
\beqn \label{milimpact}      
\fl  \int d^2\mbox{\boldmath $k$}_1 d^2\mbox{\boldmath $k$}_2 \; M_{ij} M_{ij}^*& =&       
324\;\alpha(1-\alpha)Q^2 \int d^2\mbox{\boldmath $r$}_1 d^2\mbox{\boldmath $r$}_2 \;      
K_1^2 \left (\sqrt{\alpha(1-\alpha)Q^2\mbox{\boldmath $r$}_1^2} \right )\nonumber\\       
\fl &\times&  \frac{\mbox{\boldmath       
$r$}_1^2}{\mbox{\boldmath $r$}_2^2\; (\mbox{\boldmath $r$}_2-\mbox{\boldmath       
$r$}_1)^2} \;\left[\hat{\sigma}(\mbox{\boldmath $r$}_2)+\hat{\sigma}(\mbox{\boldmath       
$r$}_2-\mbox{\boldmath $r$}_1) -\hat{\sigma}(\mbox{\boldmath $r$}_1)\right]^2 .       
\eeqn      
The vector $\mbox{\boldmath $r$}_1$ represents       
the separation of the quark and antiquark, $\mbox{\boldmath $r$}_2$       
the separation of the quark and gluon and $\mbox{\boldmath $r$}_2-\mbox{\boldmath       
$r$}_1$        
the separation of the antiquark and gluon. The combination in which      
the dipole cross section appears in (\ref{milimpact}),      
\be      
\label{eq:49}      
\hat{\sigma}(\mbox{\boldmath $r$}_2)+\hat{\sigma}(\mbox{\boldmath $r$}_2-      
\mbox{\boldmath $r$}_1)      
-\hat{\sigma}(\mbox{\boldmath $r$}_1),      
\ee      
reflects the scattering of the three effective colour dipoles, $q g$, $\bar{q} g$ and       
$q\bar{q}$, on the target proton.  Note that the scattering of the      
$q\bar{q}$-dipole in this configuration in not colour suppressed      
by powers of $N_c$. The reason for this is a contribution in which      
the initial $q\bar{q}$-pair interacts before the gluon is emitted      
(second diagram of Fig.~\ref{diag6}).      
Similar results have been found in Ref.~\cite{NZ3}. In the short distance      
limit, $r_1 \ll r_2$, the $q\bar{q}$-pair recombines effectively into a gluon,      
forming one pole of the gluon-dipole. The scattering is given by a single      
cross section, $\hat{\sigma}(\mbox{\boldmath $r$}_2)$, for the gluon-dipole.      
      
The result for the triple Regge limit is consistent with previous results when, in addition to 
the ordering of the longitudinal components, the ordering of the transverse momenta is 
assumed.  In this limit ($\beta \rightarrow 0$) one can show that (\ref{fqqg}) and 
(\ref{fqqgtr}) coincide.      
      
We have originally argued that the inclusion of a $q\bar{q}g$ component of the      
photon wave function at large $M$ is enough for most purposes. In the limit of really high       
energies, however, with each order in $\alpha_S$ one gains a logarithm in $M^2/Q^2$      
which can overcome the smallness of the coupling. The resummation of the $\alpha_S       
\log(M^2/Q^2)$ terms corresponds to the emission of multiple $s$ channel gluons. In       
Ref. \cite{BarWu} it has been shown that the $t$ channel structure then becomes  
complicated.         
The two two-gluon exchange ladders in the diagram for the diffractive structure function       
interact leading to a 4-gluon \lq\lq bound state\rq\rq~in the $t$ channel.  There is no simple       
coupling of three gluon ladders with a local triple \lq BFKL\rq~ladder       
coupling\footnote{Loosely speaking, BFKL evolution \cite{BFKL} is the resummation of       
$\alpha_S \log(1/x)$ terms applicable at small $x$, whereas DGLAP evolution       
\cite{DGLAP} corresponds to the resummation of $\alpha_S \log Q^2$ terms.} which would       
enable the two       
exchange ladders to coalesce into a single ladder.  Nonetheless, one can find special       
kinematical configurations such that a triple ladder configuration exists \cite{BarLotWu}.      
This happens at $t=0$ when the upper ladder is forced into the DGLAP regime where      
leading logs in $Q^2$ become dominant. The lower ladders on the other hand are       
of \lq BFKL' type. This result is due to a certain scaling behaviour which leads to the      
\lq conservation' of anomalous dimensions of the three ladders. For $t \ne 0$, however,      
this scale behaviour is absent and the coupling of three BFKL ladders is in principle      
possible except that the 4-gluon bound state supersedes the simple triple ladder      
scenario.      
      
\section{Multiple gluon exchange and the semiclassical approach}      
\begin{figure}[htb]      
\begin{picture}(15,3)(0,1.5)       
\put(0,0){\epsfig{figure=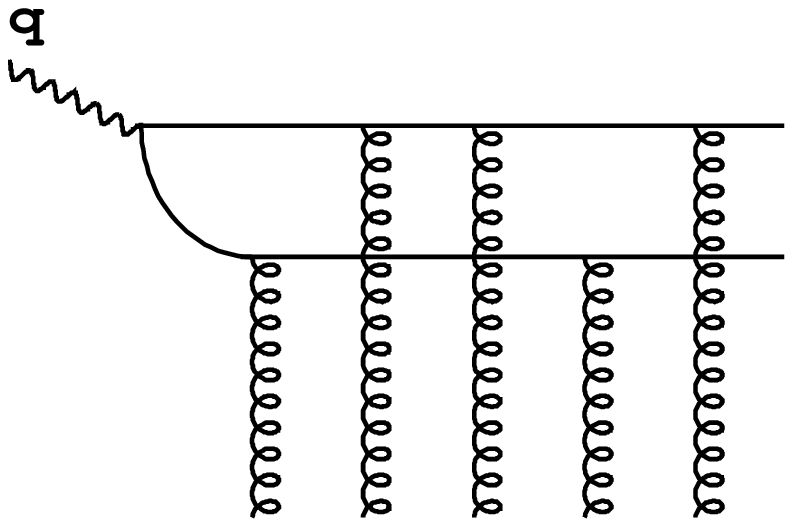,height=5.5cm}}      
\put(6.5,3){\begin{Huge}$\rightarrow$ \end{Huge}}      
\put(6.75,0){\epsfig{figure=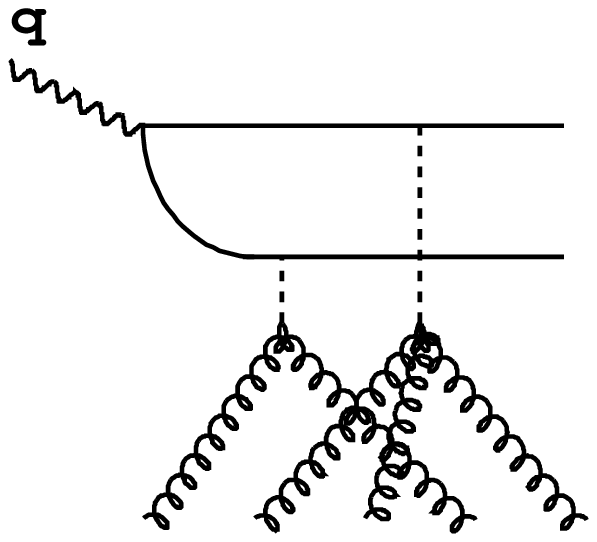,height=5.5cm}}      
\end{picture}      
\caption{\it Reduction of multiple-gluon exchange into eikonal factors shown by the dashed     
lines.  The colour structure is discussed separately.  \label{diag3}}      
\end{figure}      
So far as we have discussed diffractive processes using the perturbative assumption      
of two-gluon exchange.  Is perturbative QCD really applicable, particularly as it embraces       
potentially \lq soft\rq~contributions where $\alpha_S$ becomes large?  In the soft regime we       
would expect multiple gluon exchange. In QED the generalization from two- to       
multi-gluon exchange is fairly straightforward. One simply has to introduce eikonal factors       
\cite{BjSop} which account for the effect of resumming multiple photon      
exchanges, while the photon wave function does not change.      
One can visualize this procedure by merging any number of photons coupled      
to one arm of the dipole into one effective vertex (Fig.~\ref{diag3}). In QCD, due to       
non-commutation, the colour structure turns out to be non-trivial.  However this can be       
absorbed into the parameterization of the unintegrated gluon distribution. What matters is the 
relative colour factor of the  quark dipole compared to the gluon dipole (see Fig.~3).  Taking the limit of 
a large number $N_c$ of colours, this is the same as for two-gluon exchange.  The argument is that the       
leading-$N_c$ colour tensors, for a gluon loop with an arbitrary number of gluons attached to       
it, can be decomposed into a set of tensors with coefficients depending on $N_c$       
\cite{Carlo}.  The leading tensor is identical to the quark loop tensor times $N_c$.  The quark       
and \lq gluon\rq~loops simply come from squaring the $q\bar{q}$ and $q\bar{q}g$       
production       
amplitudes of Fig.~3.  We have illustrated this for four gluon exchange in the first row of       
Fig.~\ref{diag4}. In this scenario      
one indeed finds that multiple gluon exchange couples to a gluon or a quark dipole      
very much like a simple two-gluon exchange, i.e. all the parametrized gluonic structure      
of the \lq Pomeron' is embodied in the distribution ${\cal F}$.      
\begin{figure}[htb]      
\vspace*{-0.cm}      
\begin{picture}(15,3.5)(0,1)      
\put(2,3.5){General approach at leading $N_c$:}       
\put(1,0){\epsfig{figure=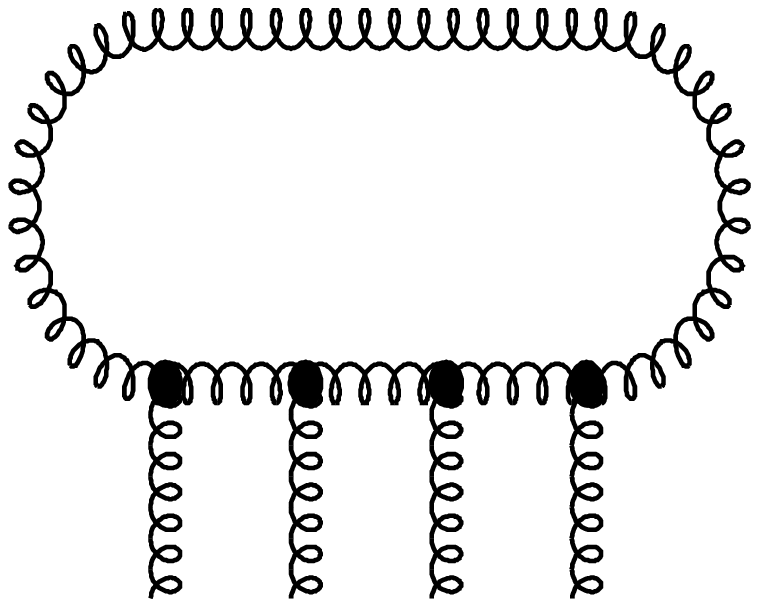,height=3.5cm}}      
\put(5,2){\begin{LARGE}$\rightarrow$ \end{LARGE}}      
\put(6.5,2){\begin{Large}N$_c$\end{Large}}      
\put(6.5,0){\epsfig{figure=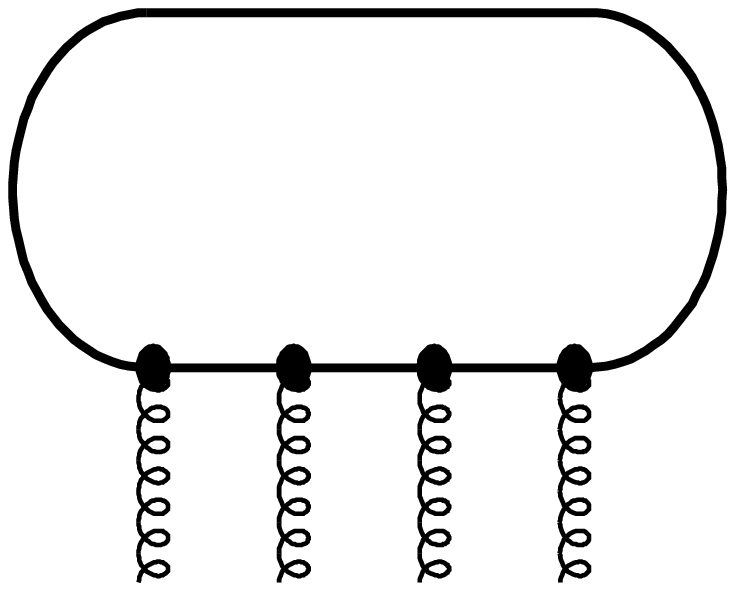,height=3.5cm}}      
\put(10.5,2){\begin{Large}+ $\cdot \cdot \cdot$\end{Large}}      
\end{picture}      
\begin{picture}(15,3.4)(0,.5)       
\put(2,3.2){'Glauber` model of non-interacting (colourless) ladders:}      
\put(1,0){\epsfig{figure=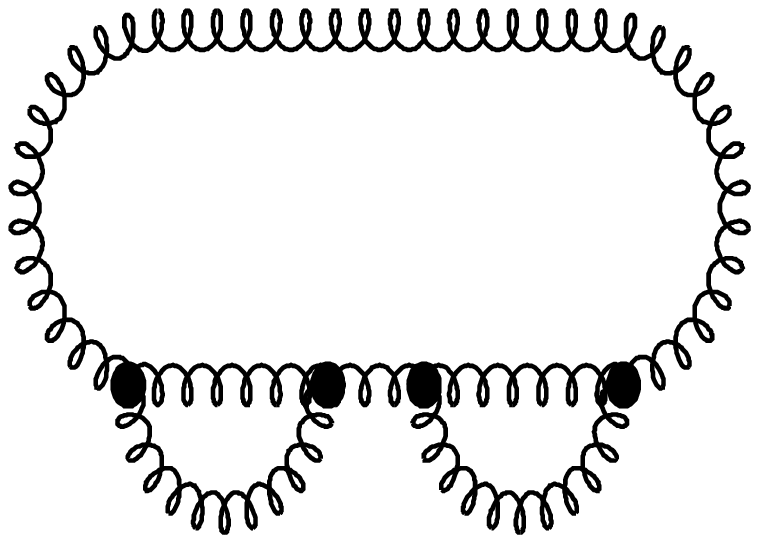,height=3.5cm}}      
\put(5,2){\begin{LARGE}$\rightarrow$ \end{LARGE}}      
\put(6.3,2){\begin{Large}4 N$_c$\end{Large}}      
\put(6.5,0){\epsfig{figure=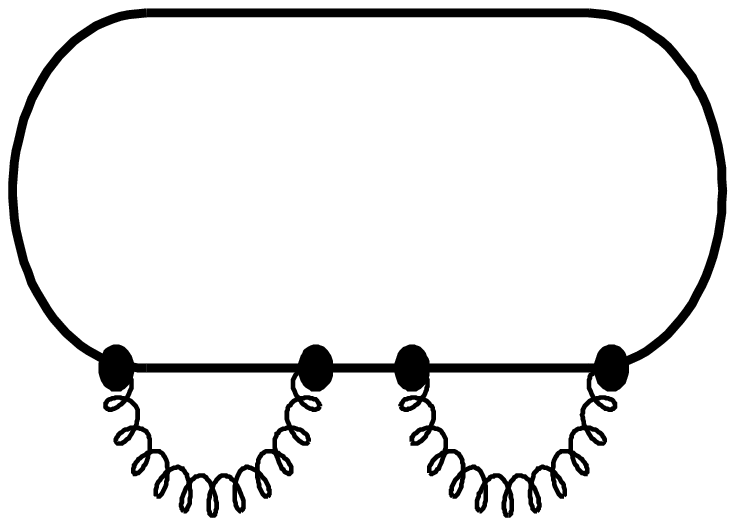,height=3.5cm}}      
\end{picture}      
\caption{\it Relating the colour structure of the cross section for $(q\bar{q})g$ production to       
that for $q\bar{q}$ production in the large $N_c$ limit for the example of four-gluon       
exchange.  In the Glauber approach the factor is $2^P N_c$ where $P$ is the number of  
ladders       
exchanged, whereas in the general case the factor is always $N_c$ in the leading $N_c$ limit.  
\label{diag4}}      
\end{figure}      
      
We may confront the general approach with a frequently used simplified model in       
which it is assumed that the gluon ladders are non-interacting.  This model is a specific subset       
of the general $t$ channel gluon structure and it is often assumed, though as yet unproven,       
that this represents the dominant configuration.  It is associated with shadowing and unitarity       
corrections and so we call it the Glauber model.  In this approach where the exchange of       
multiple {\it colourless} gluon pairs is resummed the situation looks different to the, more       
general, interacting ladder scenario.  The projection on pairs of colour singlet states mixes the      
leading $N_c$ tensor with non-leading tensors. With regard to the overall counting of powers       
of $\alpha_S N_c$ the Glauber approach is subleading. However, the only effect of summing      
the leading $\alpha_S N_c$ terms (associated with the leading-$N_c$ colour tensors mentioned earlier) 
might be the reggeization of      
the initial gluon pair and would, therefore, not contribute to unitarization or shadowing.      
In this case the Glauber type multiple scattering would gain importance.            
As the second row of Fig.~\ref{diag4} shows, once the four      
gluons are projected pairwise on a colour singlet state the       
relative colour factor of a gluon loop compared to a quark loop      
is $4 N_c$ instead of $N_c$. The factor $4$ is the large $N_c$ limit of the ratio      
$C_A^2/C_F^2$, i.e. for each pair of gluons we find a relative factor $C_A/C_F$       
for the scattering of a gluon dipole as compared to a quark dipole. In this scenario the      
parameterization for the unintegrated gluon distribution $\cal{F}$ would be different      
for $q\bar{q}$ and $q\bar{q}g$ production. Both contributions would no longer be related      
as in (\ref{fqqt}) and (\ref{fqqg}).      
      
The semiclassical approach in Refs.~\cite{BH,BGH} is conceived as a       
non-perturbative approach (for a recent review see Ref.~\cite{Heb}) and      
utilizes Wilson loops. It is very much reminiscent of the eikonal      
approach in QED. According to our preceding arguments the photon wave functions      
turn out to be the same as in the two-gluon exchange. The differences      
emerge in modeling the dipole cross section. The semiclassical model shows similarities      
to the Glauber approach and has therefore not quite the structure of (\ref{fqqt}) and     
(\ref{fqqg}).      
The difference occurs in the treatment of the colour as we have discussed above.      
In contrast to the Glauber approach the energy dependence is introduced as part       
of the overall normalization, which is taken to have the      
form $\Omega (L-\ln x)^2$, where $\Omega$ and $L$ are parameters to be determined by the  
data. It is argued that this form of energy dependence is      
due to the number of field modes which increases when      
the energy is increased. It is consistent with unitarity      
constraints.      
      
\section{Colour dipole approach}      
The colour dipole formalism has been developed in \cite{MuePat,NNCD,NZ3} as an       
alternative to the Feynman diagram approach to small $x$ physics. It is formulated      
in impact parameter space and has been shown      
to reproduce Feynman diagram results for inclusive processes in the Regge limit, as embodied in the     
BFKL equation (which will be introduced in Section 13).      
With regard to gluon radiation in diffraction it can be applied in the triple Regge limit,      
i.e. for large masses $M^2 \gg Q^2$ only.      
At lowest order and first non-leading order in the strong coupling,       
i.e. $q\bar{q}$ and $q\bar{q}g$ final states, the colour dipole approach      
is consistent with the results derived in Section 7. In particular it can be proven to coincide       
with the triple Regge result of (\ref{milimpact}) \cite{NZ3,Mueprivate}.      
One expects that multiple gluon radiation in the colour dipole approach      
leads to the same results as the Feynman diagram approach \cite{BarWu}.      
A complete phenomenological treatment of diffraction within the framework      
of the colour dipole approach has been presented in Ref.~\cite{BPR}.      
The colour dipole approach of Ref.~\cite{BPR} uses the non-forward       
BFKL-Pomeron and, in this respect, goes beyond other approaches. It       
directly delivers the $t$ integrated cross section.       
      
\section{Diffractive production of open charm in DIS}      
The measurement of charm in diffractive scattering provides       
an additional test for any approach based on the photon wave      
function formalism. As before in the case of light flavours we consider      
the exclusive $c\bar{c}$-pair which arises from the dissociation of       
longitudinally and transversely polarized photons, as well as the production of the       
$c\bar{c}g$-state.      
Since the mass of the charm quark sets a limit on the size of the $c\bar{c}$-dipole,      
it becomes colour transparent and one expects       
a strong suppression for this configuration.      
The effective gluon dipole, associated with $c\bar{c}g$ production,       
on the other hand, is not restricted in size.      
      
For a quantitative analysis we have to extend our expressions for      
$q\bar{q}$ production to include the charm mass $m_c$.       
We find that the diffractive structure functions for $\gamma^* p \rightarrow c\bar{c}p$ are       
\cite{LMRT,Lot,Diehlcc}      
\beqn       
\fl & & x_{\fP}F_{T,c\bar{c}}^D(\beta, Q^2, x_{\fP}) =  \frac{e_c^2}{48 B_D}\;       
\frac{\beta}{(1-\beta)^2}\int \frac{dk_t^2}{k_t^2} \;\frac{k_t^2+m_c^2}{\sqrt{1-4\beta       
k^2/Q^2}}\nonumber\\      
\fl & & \quad\quad\quad\quad\quad\quad \times \Theta\left (k^2-\frac{Q^2}{4\beta}\right) \:       
\left\{\left[1-\frac{2\beta k^2}{Q^2}\right] \;|I_T|^2 + \frac{4       
k_t^2 m_c^2}{k^4}\;|I_L|^2\right\}, \label{eq:52} \\       
\fl & & x_{\fP}F_{L,c\bar{c}}^D(\beta, Q^2, x_{\fP})  =  \frac{e_c^2}{3 B_D Q^2} \;      
\int \frac{dk_t^2}{1-\beta}\; \frac{k^2\;\beta^3}{\sqrt{1-4\beta k^2/Q^2}} \: \Theta\left(k^2-      
\frac{Q^2}{4\beta}\right)\;\;|I_L|^2, \label{fcct}      
\eeqn      
with      
\beqn      
\fl & & I_T  =  \int\frac{d\ell_t^2}{\ell_t^2} \alpha_S{\cal F}(x_{\fP},\ell_t^2)      
\left[1-2\beta-2\frac{m_c^2}{k^2}+ \frac{\ell_t^2 - (1-2\beta)\;k^2+2\;m_c^2}{\sqrt{(\ell_t^2       
+ k^2)^2 -4 \ell_t^2 \;k_t^2}} \right], \label{it} \\      
\fl & & I_L  =  \int\frac{d\ell_t^2}{\ell_t^2 
} \alpha_S{\cal F}(x_{\fP},\ell_t^2)      
\left[1\;-\;\frac{k^2} {\sqrt{(\ell_t^2+k^2)^2 -4 \ell_t^2\;k_t^2}} \right]\;\;. \label{il}      
\eeqn      
The definition of $\beta$ is the same as before, i.e. $\beta=Q^2/(M^2+Q^2)$, but      
the diffractive mass $M$ has now a lower limit given by $2 m_c$.  Hence the $\beta$       
spectrum has an upper limit well below 1 at the lower values of $Q^2$.  Also the      
two-body kinematical relation (\ref{kin1}) acquires a mass term and changes into       
\beqn      
\label{eq:a57}    
k^2&=&\frac{k_t^2+m_c^2}{1-\beta}=\alpha(1-\alpha)\;\frac{Q^2}{\beta}\;\;.      
\eeqn      
Note the second term in the curly brackets of (\ref{eq:52}) which      
results from a spin flip on the quark line. The spin flip is only present in the       
massive case.      
      
We may go a step further and investigate the limit of small $\ell_t^2$ in the expressions      
(\ref{it}) and (\ref{il}). The relevant scale for the expansion is $k^2$, i.e. we       
expand in powers of $\ell_t^2/k^2$:      
\beqn      
I_T&\approx&\label{itapproax}      
\int^{k^2}\frac{d\ell_t^2}{\ell_t^2} \alpha_S{\cal F}(x_{\fP},\ell_t^2)\;      
\left[1+\left(1-2\frac{k_t^2}{k^2}\right)\left(1-2\beta-2\frac{m_c^2}{k^2}\right)      
\right]\;\frac{\ell_t^2}{k^2}\nonumber \\      
&\approx& \frac{ \alpha_S x_{\fP} g(x_{\fP},k^2)}{k^2}\;      
\left[1+\left(1-2\frac{k_t^2}{k^2}\right)\left(1-2\beta-2\frac{m_c^2}{k^2}\right)      
\right], \\      
I_L&\approx&\label{ilapprox}      
\int\frac{d\ell_t^2}{\ell_t^2} \alpha_S{\cal F}(x_{\fP},\ell_t^2)\;      
\left(1-2\frac{k_t^2}{k^2}\right)\;\frac{\ell_t^2}{k^2}\nonumber\\      
&\approx& \frac{ \alpha_S x_{\fP} g(x_{\fP},k^2)}{k^2}\;      
\left(1-2\frac{k_t^2}{k^2}\right)\;\;.      
\eeqn      
Here $g$ is the conventional gluon distribution of the proton, obtained by integrating the       
unintegrated distribution $\cal F$.      
The interesting point about the scale $k^2$ is its lower cut-off $m_c^2/(1-\beta)$      
as $k_t$ approaches zero. This means that the diffractive production of exclusive      
$c\bar{c}$-pairs always stays in the hard or perturbative regime regardless of the value for       
$k_t$.      
At large $k_t^2$, i.e. $k_t^2 \gg m_c^2$, the distribution in $k_t^2$ falls      
roughly as $1/k_t^4$ for the transverse part.       
We can therefore approximate the $k_t^2$-integration for the transverse part by      
replacing the argument $k^2$ in the structure function by $m_c^2/(1-\beta)$ and      
perform the integral for the remaining terms (see also Ref.~\cite{GNZ}):      
\beqn      
x_{\fP}F_{T,c\bar{c}}^D(\beta, Q^2, x_{\fP}) &\approx&      
 \frac{e_c^2}{72 B_D m_c^2}\;\beta\;(1-\beta)^2\;(3+4\beta+8\beta^2)\nonumber\\      
&\times&\left[\alpha_S x_{\fP} g\left(x_{\fP},\frac{m_c^2}{1-\beta}\right)\right]^2      
\;\;.      
\eeqn      
The cross section shows a rapid growth proportional to $(x_{\fP} g)^2$ as long       
as the saturation      
regime is not reached. The prefactor $1/m_c^2$, on the other hand, leads to a       
suppression as compared to light flavours (colour transparency).      
The longitudinal part has a $k_t^2$-spectrum which is proportional to $1/k_t^2$      
so that a logarithm builds up after integrating over $k_t^2$:       
\beqn      
x_{\fP}F_{L,c\bar{c}}^D (\beta, Q^2, x_{\fP}) &\approx&      
 \frac{e_c^2}{3 B_D Q^2}\;\beta^3 \;(1-2\beta)^2      
\;\ln\left(\frac{(1-\beta)\;Q^2}{4\beta\; m_c^2}\right)\nonumber\\      
&\times&\left[\alpha_S x_{\fP} g\left(x_{\fP},\frac{Q^2}{4 \beta}\right)\right]^2      
\;\;.      
\eeqn      
It has already been noted that the longitudinal part is a genuine hard      
contribution which can be calculated using (\ref{l-second}). The price      
to pay is the factor $1/Q^2$ which makes it a higher twist contribution. The dependence of 
the unintegrated gluon distribution ${\cal F}$ on $k_t^2$ cannot be completely 
neglected. In this sense the previous equation represents only a crude estimate.      
      
For the $c\bar{c}g$ component we will proceed in a slightly different way,       
making use of the diffractive factorization property. In (\ref{gluon-distr})      
we have specified the diffractive gluon distribution which needs to be folded with      
the corresponding charm-coefficient function \cite{WITTEN,LMRT}:      
\beqn      
F_{c\bar{c}g}^D (\beta, Q^2, x_{\fP}) \label{fccg}      
&=&      
2 \;\beta\; e_c^2\; \frac{\alpha_S}{4 \pi}\;      
\int_{a \beta}^1 \frac{dz}{z}\;C_2\left(\frac{\beta}{z},\frac{m_c^2}{Q^2}\right)      
\; g^D(z)      
\eeqn      
where $a=1+4m_c^2/Q^2$ and      
\beqn      
C_2(y,r)&=&\left[y^2+(1-y)^2+4y(1-3y)r-8y^2r^2\right]\;      
\ln\frac{1+v}{1-v}\nonumber\\      
&&+v\left[-1+8y(1-y)-4y(1-y)r\right]\;\;.      
\eeqn      
$v$, the centre-of-mass velocity of the charm quark or antiquark, is given by      
\be      
v^2\;=\;1-\frac{4ry}{1-y}\;\;.      
\ee      
      
In Ref.~\cite{LMRT} the diffractive gluon distribution was calculated      
assuming a cut-off $k_0^2$ on $k_t^2$ (see also Ref.~\cite{LeWu}). This result can       
be rederived from the momentum representation\footnote{The impact representation of       
$g^D (\beta)$ was given in (\ref{gluon-distr}).} of $g^D$ by expanding in powers      
of $\ell_t^2/k^2$:      
\beqn      
\label{gluondistr2}      
\fl & & g^D(\beta) = \frac{9}{64 x_\funp B_D}\;\frac{1}{\beta\,(1-\beta)}\;       
\int_{k_0^2}^{\infty} dk_t^2\;\left\{\int\frac{d\ell_t^2}{\ell_t^2}\;\alpha_S      
{\cal F}(x_{\fP},\ell_t^2) \;\; \right. \nonumber \\      
\fl && \quad\quad\quad\quad\quad \times \left.\left[\beta^2+(1-\beta)^2+\frac{\ell_t^2}{k^2}-      
\frac{[(1-2\beta)k^2-\ell_t^2]^2+2\beta(1-\beta)k^4}      
{k^2\sqrt{(\ell_t^2+k^2)^2-4(1-\beta)\;\ell_t^2\;k^2}}      
\right]\right\}^2 \nonumber\\      
\fl & & \quad\quad\quad \approx \frac{9}{16 B_D} \;\frac{1}{\beta}\;(1-      
\beta)^3\;(1+2\beta)^2\; \int_{k_0^2}^{\infty}\frac{dk_t^2}{k_t^4}\; \left[ \alpha_S x_{\fP}       
g(x_{\fP},k^2)\right]^2      
\eeqn      
bearing in mind that $k^2=k_t^2/(1-\beta)$. Another way of deriving the second      
equality would be the use of the second derivative (\ref{g-second}) of the $\Psi_{q\bar{q}g}$       
wave function. Unfortunately only the hard regime is being correctly taken into account.       
More appropriate is the use of the full expression in the first line of (\ref{gluondistr2}).
      
Another comment is appropriate, which concerns the possible $K$ factors suggested in       
Ref.~\cite{LMRT}.  $K$ factors are used to represent the higher order corrections to the       
process.  A certain virtual correction has been singled out which in an extended analogy      
to inclusive Drell-Yan pair production might lead to a significant enhancement of the       
prediction for diffractive scattering in general, not only for diffractive charm production. In       
order to understand the importance of higher order corrections a complete NLO calculation is       
required.      
      
The particular signature for diffractive charm production is a      
fairly large contribution around 25\% at small $\beta$ where the $c\bar{c}g$      
component dominates. The fraction of charm in this regime is the same as      
expected in inclusive charm production. At larger $\beta$ and large $Q^2$, i.e. $\beta>0.2$ and 
$Q^2 > 10$~GeV$^2$, however, charm is suppressed due to colour transparency. It contributes      
only a few percent, with a possible $K$ factor enhancement. This feature helps to       
discriminate against other approaches (for example in Ref.~\cite{H1diff})       
which suggest a hard and dominant gluon distribution at large $\beta$      
and consequently a much larger charm fraction.      
      
\section{Diffractive vector meson production}      
The diffractive electroproduction of vector mesons, $\gamma^* p \rightarrow Vp$ offers the       
opportunity to study many aspects of perturbative QCD, as well as revealing information on       
the gluonic structure of the proton. The hard scale can be either the virtuality $Q^2$ of the       
photon \cite{COLLFS}, the mass of the quarks for heavy vector mesons, the momentum transfer $t$ or some       
combination.  Data are available for $\rho, \omega, \phi, J/\psi, \psi^\prime, \Upsilon$ vector       
meson production.  Moreover the observed decays such as $\rho \rightarrow \pi^+ \pi^-$ and       
$J/\psi \rightarrow \mu^+ \mu^-$ allow a helicity decomposition of the $\gamma^* p       
\rightarrow Vp$ amplitudes.  The basic mechanism is the same as that for open $q\bar{q}$       
production, except that we must now include a wave function $\psi^V (q\bar{q})$ for the       
eventual formation of the vector meson, see Fig.~8.  Since, at high energies, the $\gamma^*       
\rightarrow q\bar{q}$ fluctuation time and the $q\bar{q} \rightarrow V$ formation time are       
both much longer than the $q\bar{q}$ interaction time with the proton, the amplitude has the       
factorized structure      
\be      
\label{eq:A}      
A (\gamma^* p \rightarrow Vp) \; = \; \psi_{q\bar{q}}^\gamma \: \otimes \: A_{q\bar{q} +       
p} \: \otimes \: \psi_{q\bar{q}}^V.      
\ee      
      
We must also note that $\gamma^* p \rightarrow Vp$ is described by \lq\lq       
skewed\rq\rq~parton distributions of the proton, that is $x, \mbox{\boldmath $\ell$}_t \neq       
x^\prime, \mbox{\boldmath $\ell$}_t^\prime$ in Fig.~8.  Reviews\footnote{There is much recent activity, and 
a large literature, concerning skewed parton distributions.  For the high energy diffractive processes that 
we discuss, we are interested in distributions in the small $x$ domain.  A general review focusing on their 
use in virtual Compton scattering can be found in \cite{GV}.} of \lq\lq       
skewed\rq\rq~(also called off-forward, non-forward or off-diagonal) distributions can be       
found in \cite{Ji,RAD,MR,GM}.  For $t = 0$ the skewedness comes from $x \neq x^\prime$,       
and increases as the mass of the vector meson $M_V$ or the photon virtuality increase.        
However, for small $x$, relevant to $\gamma^* p \rightarrow Vp$ at high energies, the       
skewed distributions are completely determined by the conventional diagonal distributions       
\cite{SHUV}.  Since the cross sections depends on the square of the gluon       
distribution, these processes are expected to be a particularly sensitive constraint on the       
gluon. The subtlety is the determination of the vector meson wave      
function. Various approaches give quite different results.      
The corresponding uncertainty mainly affects the absolute normalization, and      
less the energy dependence of the cross section. But even the       
energy dependence is not perfectly well established, for      
the scale which enters the structure function is not linked to the hard scale in a simple way.       
\begin{figure}[htb]   
 \begin{picture}(15,4.5)(0,1)    
\put(2.5,0.2){\epsfig{figure=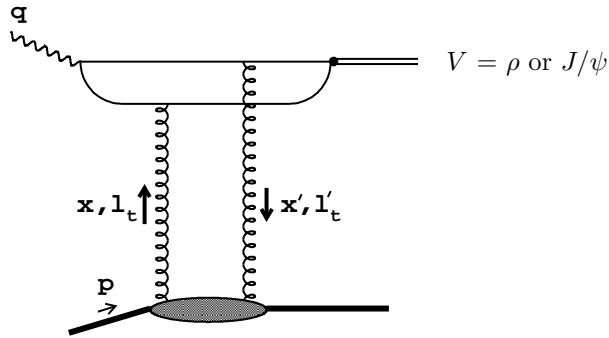,height=6cm}}    
\put(9.5,4.7){$V$ = $\rho$ or $J/\psi$}  
  \end{picture}       
\caption{\it One of the four basic diagrams for diffractive vector meson electroproduction,       
$\gamma^* p \rightarrow Vp$.  (There are three other couplings of the gluons to the       
$q\bar{q}$ pair.)  In general $x \neq x^\prime$, $\ell_t \neq \ell_t^\prime$ and the cross     
section depends on the square of the \lq\lq  skewed\rq\rq~gluon distribution.  \label{diag7}}      
\end{figure}      
      
On the other hand for large $t$ the skewedness in the gluon ladder comes from $\mbox{\boldmath $\ell$}_t \neq       
\mbox{\boldmath $\ell$}_t^\prime$ and the evolution follows BFKL.  Large $t$ diffraction      
is an ideal place to study the high energy limit of pQCD \cite{ForRys}. Unfortunately large 
$t$ vector meson production is again hampered with the problem of how well the meson 
wave function is understood \cite{BFLW}. It has therefore been proposed to look for photons      
instead of vector mesons in the final state \cite{IvWu,GI}, i.e. quasi elastic      
$\gamma p$-scattering. At $t=0$, but large $Q^2$, this would be Deeply Virtual Compton 
Scattering (DVCS) \cite{FFS}.    
      
Vector meson production has developed into a very active field which deserves a separate       
review.  Here we will concentrate on the perturbative QCD description of these processes,  
but note that there are also approaches which are based on the vector meson dominance   
(VMD) and related models \cite{FRS,DL-rho}.  First we will study $J/\psi$ and $\rho$   
production at $t = 0$ and then, in the following section, $t \neq 0$ production.

\section{Diffractive $J/\psi$ production at $t=0$}      
We start our discussion with the electroproduction of $J/\psi$ \cite{BJ,RysVM}.       
The modeling of the $J/\psi$ wave function is in general a non-trivial procedure  
\cite{BFGMS,FKS,LMRR} but, for the purposes of illustration, we use       
the nonrelativistic, static wave function assumption made in Refs.~\cite{BJ,RysVM}.      
      
The production mechanism for vector mesons is closely related      
to open diffraction. The photon dissociates into a $q\bar{q}$-pair which then       
scatters off the target. With a certain probability       
the final state recombines into a      
vector meson. This probability is determined by the projection on the      
meson wave function. The photon dissociation is as before described      
by the photon wave functions (\ref{eq:g15}-\ref{eq:h15}).       
We can make a simple qualitative estimate about the longitudinal versus      
the transverse cross section assuming that $Q^2$ is much bigger than      
$M_{\psi}^2$ ($M_{\psi}$ is the mass of the $J/\psi$)       
and the corresponding $\beta$      
is close to 1. Whereas the transverse component\footnote{Note that the expression in the 
curly brackets in (\ref{fqqt}) vanishes as $\beta \rightarrow 1$, and overcomes the prefactor 
$1/(1- \beta)$.}      
is decreasing when $\beta$ is approaching 1 the longitudinal component      
remains constant. This already indicates the dominance      
of the longitudinal versus the transverse cross section.      
      
The wave function of a vector meson is constructed in analogy to      
the photon wave function. In the simplified nonrelativistic, static case one      
assumes that the quark and antiquark are weakly bound in comparison to their mass      
and that, therefore, their relative movement is small.       
This means that the relative transverse momentum      
is approximately zero, i.e. $k_t\approx 0$ and       
that the quark and antiquark equally share the longitudinal momentum fraction,      
i.e. $\alpha=1/2$. It also means that the mass of the $J/\psi$      
is approximately the sum of the mass of charm and anticharm quark, i.e.       
$M_{\psi}=2m_c$. In this simplified approach one can write      
\be\label{psiv}      
\Psi^V(\mbox{\boldmath $k$}_t,\alpha)\;\sim\;\delta(\alpha-1/2)\;\delta^2(\mbox{\boldmath       
$k$}_t)\;\;.      
\ee      
We take the expressions for open charm production,      
(\ref{eq:52}) and (\ref{fcct}), and apply      
the $\delta$-functions of (\ref{psiv}). To this end one has       
to integrate over $\beta$ using the relation (\ref{eq:a57}) which      
connects $\alpha$ with $\beta$. The first thing to notice when      
$k_t$ is set to zero is that the integral $I_T$ of (\ref{it}) vanishes. The final state  
configuration      
has to match the spin of the incoming photon. For a transverse      
polarized photon this means that the final state has to have the      
total angular momentum $J_z=\pm 1$. This can be the \lq orbital'       
angular momentum of the emerging charm-anticharm pair, $L_z=\pm 1$,      
when the quark spins are opposite, i.e. $S_z=0$. The other possibility is       
spin-flip along the quark line, $S_z =\pm 1$, while the orbital angular       
momentum $L_z=0$. For the orbital      
angular momentum to be non-zero the transverse momentum $k_t$ has to be      
non-zero. Since we require $k_t=0$ we enforce $L_z=0$ and $I_T = 0$, leading to the     
vanishing of the first term in the curly bracket of (\ref{eq:52}). The second      
term, which is associated with the spin-flip, remains and leads, after a little algebra, to:      
\beqn \label{t-psi}      
x_{\fP}F_{T,\psi}^D(Q^2, x_\funp) \label{fVt}      
& \sim & \frac{Q^4 M_{\psi}^2}{(Q^2+M_{\psi}^2)^3}      
\;\left\{\int d\ell_t^2\;\frac{\alpha_S{\cal F}(x_{\fP},\ell_t^2)}      
{\ell_t^2+ \bar{Q}^2}\right\}^2 \nonumber \\      
& \approx  & \frac{16\;Q^4       
M_{\psi}^2}{(Q^2+M_{\psi}^2)^5}      
\;\left[ \alpha_S x_{\fP} g (x_{\fP}, \bar{Q}^2) \right]^2      
\eeqn      
where $x_{\fP}=(Q^2+M_{\psi}^2)/(Q^2+W^2)$, see (\ref{eq:a3}) and (\ref{eq:a5}), and       
\be      
\label{eq:z1}      
\bar{Q}^2 \; = \; \alpha (1 - \alpha) Q^2 + m_c^2 \: \simeq \: (Q^2 + M_\psi^2)/4.      
\ee      
The leading contribution to the $\ell_t^2$ integration comes from the region $\ell_t^2<       
\bar{Q}^2$.      
      
For longitudinal polarized photons both the orbital angular momentum and the spin are zero,       
i.e. $L_z=0$ and $S_z=0$      
(no spin-flip). There is no suppression when $k_t$ is set to zero      
\beqn \label{l-psi}      
x_{\fP}F_{L,\psi}^D(Q^2,x_\funp) \label{fVl}    
& \sim & \frac{Q^6}{(Q^2+M_{\psi}^2)^3}      
\;\left\{\int d\ell_t^2\;\frac{\alpha_S{\cal F}(x_{\fP},\ell_t^2)}      
{\ell_t^2 + \bar{Q}^2} \right\}^2 \nonumber \\      
& \approx & \frac{16\;Q^6}{(Q^2+M_{\psi}^2)^5}      
\;\left[ \alpha_S x_{\fP} g\left(x_{\fP}, \bar{Q}^2 \right)\right]^2.      
\eeqn      
We see that, in this simple approach, the transverse and the longitudinal production      
of $J/\psi$ mesons are identical up to a factor $Q^2/M_{\psi}^2$ by which      
the longitudinal is enhanced over the transverse component. Also       
the dependence on the square of the gluon distribution is evident.      
      
With regard to the absolute normalization one has to go further into the details      
of the construction of the vector meson wave function. For reasons of consistency      
the same wave function determines the leptonic decay width for the $J/\psi$       
\cite{BJ,BFGMS}.  This allows the normalization to be determined by the decay width       
$\Gamma_{e^+e^-}^{\psi}$.  The complete expression for the cross section (longitudinal and       
transverse) then reads \cite{RysVM,LMRR}:      
\beqn \label{psi-x}      
\fl \left.\frac{d\sigma^{\psi}}{dt}\right|_{t=0}&=&      
\frac{16\;\Gamma^{\psi}_{e^+e^-}\;M_{\psi}^3 \;\pi^3}{3\;       
\alpha_{em}\;(Q^2+M_{\psi}^2)^4}\;      
\left[ \alpha_S x_{\fP} g\left(x_{\fP}, \bar{Q}^2 \right)\right]^2 \: \left ( 1 +  
\frac{Q^2}{M_\psi^2} \right ).      
\eeqn      
      
The relativistic corrections to the $J/\psi$ wave function (\ref{psiv}) have been considered in  
Refs.~\cite{FKS,LMRR,HOODB}.  The change that these make to the cross section is under  
dispute.  Refs.~\cite{LMRR,HOODB} show that the net effect of the corrections is small.   
On the other hand the authors of Ref.~\cite{FKS} argue that they cause a large suppression.   
These corrections, together with the estimates of the higher order contributions, lead to an  
uncertainty in the absolute normalization\footnote{The use of a skewed gluon distribution  
causes a 30\% enhancement of the cross section for $J/\psi$ photoproduction \cite{SHUV}.},  
but have much less impact on the predictions for the dependence of the cross section on the       
$\gamma^* p$ energy $W$.     
  
\begin{figure}[htb]        
  \vspace*{-0.5cm}       
\begin{center}      
         \epsfig{figure=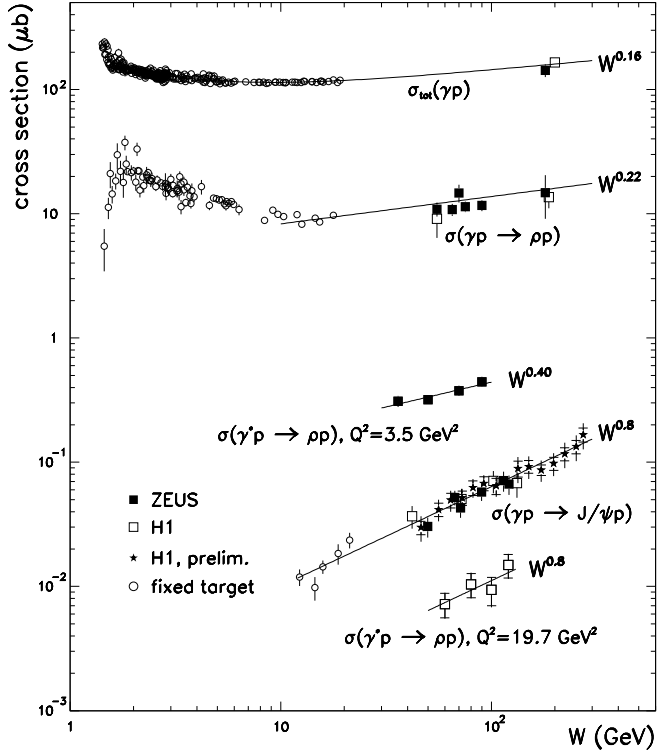,height=11cm}       
\end{center}       
\vspace*{-0.cm}       
\caption{\it The energy dependence of the cross sections for diffractive $\rho$   
electroproduction (at three values of $Q^2$) and $J/\psi$ photoproduction.  Also shown is the   
total $\gamma p$ cross section compared to the $W^{0.16}$ energy dependence of  
(\ref{eq:b10}).  The $W^{0.22}$ and $W^{0.8}$ lines indicate the \lq   
soft\rq~Regge Pomeron and \lq hard\rq~QCD Pomeron predictions of (\ref{eq:70b}) and  
(\ref{eq:70a}) respectively. (A. Levy, private communication.) 
\label{vm}}       
\end{figure}       
In the photoproduction limit we simply set $Q^2=0$.      
Only the transverse component survives in this case. The large charm mass      
provides a hard enough scale for perturbation theory to be applicable.      
The experimental advantage of photoproduction is the high      
statistics compared to electroproduction. The analysis of Ref.~\cite{LMRR}      
demonstrates the value of $J/\psi$ photoproduction in determining the      
gluon distribution of the proton. Although corrections which go beyond leading      
log in $Q^2$ have an influence on the cross section, these       
mainly affect the absolute normalization. The shape of the $W$ distribution      
seems invariant under these corrections.  The prediction is      
\begin{eqnarray}   
\sigma^\psi & \sim & \left [x_\funp \: g(x_\funp, \bar{Q}^2) \right ]^2       
\nonumber \\      
& \sim & [x_\funp^{- \lambda}]^2 \; \sim \; W^{4 \lambda} \; \sim \; W^{0.8},      
\label{eq:70a}     
\end{eqnarray}  
using the rise of the gluon distribution with decreasing $x$ determined by analysis of all       
available deep inelastic scattering data.  This should be compared with the \lq soft\rq~or       
Pomeron Regge exchange prediction      
\be      
\label{eq:70b}      
\sigma^V \;\; \sim \;\; \langle (W^2)^{2 (\alpha_\funp (t) - 1)} \rangle \;\; \sim \;\; W^{0.22}      
\ee      
where the Pomeron trajectory $\alpha_\funp$ is given by (\ref{eq:a9}) and $\langle \cdots 
\rangle$ indicate that the $\gamma^* p \rightarrow Vp$ cross section is averaged over $t$.  
We see, from Fig.~\ref{vm}, that the measured energy dependence of $J/\psi$ production is 
in good agreement with (\ref{eq:70a}), which provides support for the     
perturbative QCD approach.     
 
\section{Diffractive $\rho$ electroproduction at $t = 0$} 
 
The diffractive production of $\rho$ mesons, $\gamma^* p \rightarrow \rho p$, is special  
because this is the process for which the data are most extensive at present.  However, unlike  
$J/\psi$ production, we do not have the mass of the charm quark to provide a hard scale.  The  
light $u$ and $d$ flavours do not provide the hard scale and so only electroproduction can be  
considered for a perturbative QCD treatment, where large $Q^2$ provides the scale.  In fact  
the $\gamma^* p \rightarrow \rho p$ data in Fig.~\ref{vm} show the transition from the \lq  
soft\rq~Regge prediction of (\ref{eq:70b}) at $Q^2 = 0$ to the \lq hard\rq~QCD  
prediction of (\ref{eq:70a}) at $Q^2 \approx 20$~GeV$^2$.  The other evidence that  
perturbation theory is applicable at large $Q^2$ comes from the measurements of the slope  
$B_\rho$ which defines the diffractive peak, $d \sigma/dt \sim \exp (-B_\rho |t| )$.  The value  
decreases rapidly from $B_\rho \sim 10$~GeV$^2$ at $Q^2 = 0$ to $B_\rho  
\sim$~5--6~GeV$^2$ for $Q^2 \sim 10$~GeV$^2$.  The latter value is expected from the  
size of the proton which suggests that the size of the $\gamma^* \rightarrow \rho$ vertex is  
close to zero. 

\indent From (\ref{psi-x}) we see that the leading order prediction for electroproduction in  
longitudinally polarized states is 
\be 
\label{eq:74a} 
\sigma_L \; \sim \; \frac{[xg (x, Q^2/4)]^{2}}{Q^6} \; \sim \;  
\frac{(Q^2)^{2 \gamma}}{Q^6} \; \sim \; \frac{1}{Q^{4.8}} 
\ee 
for $Q^2 \gg M_\rho^2$, where $x = Q^2/W^2$ and $\gamma$ is the effective anomalous dimension  
of the gluon density 
\be 
xg (x, Q^2) \; \sim \; (Q^2)^\gamma. 
\label{eq:74b} 
\ee 
We have taken a representative value $\gamma = 0.3$ determined from the global parton  
analyses, corresponding to the $x$ range ($10^{-3} \lapproxeq x \lapproxeq 10^{-2}$) of the  
HERA data.  The QCD prediction (\ref{eq:74a}) for $\sigma_L$ is consistent with the $Q^2$  
behaviour seen in the data.  This is not the case for $\sigma_T$.  The non-relativistic  
prediction  
for $\sigma_T$, obtained from (\ref{t-psi}) and (\ref{l-psi}), 
\be 
\label{eq:74c} 
\sigma_T \; = \; \frac{M_\rho^2}{Q^2} \;\sigma_L, 
\ee 
appears to be too small and to fall too rapidly with increasing $Q^2$, as is evident by  
comparing the dashed line in Fig.~\ref{rholt} with the data.  The problem lies with the wave  
function which describes the recombination of a $q\bar{q}$ pair into the $\rho$ meson \cite{BFGMS}.   
Unlike $J/\psi$ production, it is crucial to include relativistic, non-leading twist corrections  
even to obtain a rough estimate of $\sigma_L/\sigma_T$ \cite{MRT}.  However these turn  
out to be insufficient to resolve the problem.  Moreover there are indications \cite{MRT} that a 
non-perturbative approach will make $\sigma_T$ fall off even faster with $Q^2$ and make matters worse. 
\begin{figure}[htb]        
\vspace*{-0.5cm}       
\begin{center}      
\epsfig{figure=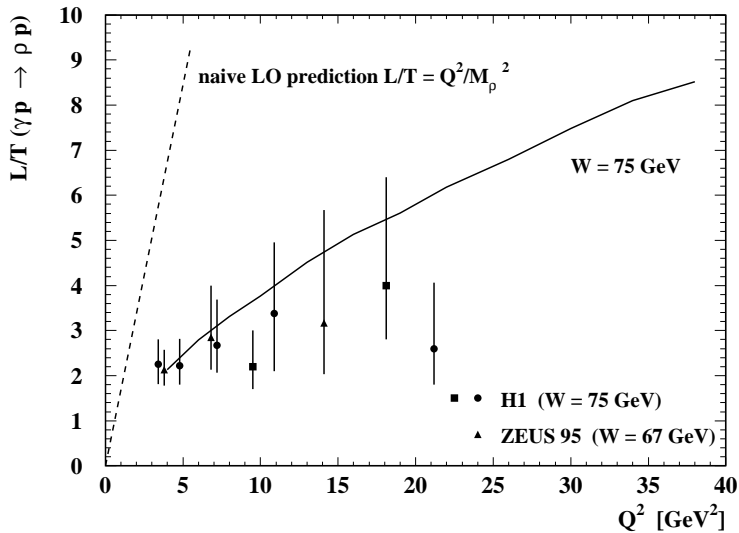,height=9cm}  
\end{center}       
\vspace*{-0.5cm}       
\caption{\it The $Q^2$ dependence of the QCD predictions \protect{\cite{MRT4}} for the  
ratio $\sigma_L/\sigma_T$ of the electroproduction of $\rho$ mesons $(\gamma^* p  
\rightarrow \rho p)$ in longitudinal and transverse polarisation states compared with HERA  
data.  The dashed line indicates the expectations of (\ref{eq:74c}).} 
\label{rholt}      
\end{figure}   
      
In order to circumvent the difficulties of modelling the $\rho$ meson wave function an  
alternative approach has been proposed \cite{MRT}.  It is based on the production of  
$u\bar{u}$ and $d\bar{d}$ pairs in a broad mass interval containing the $\rho$ resonance.  In  
this mass interval, phase space forces the $q\bar{q}$ pairs to hadronize dominantly into  
$2\pi$ states.  Moreover provided the $q\bar{q}$-proton interaction does not distort the spin,  
we expect that the transition $\gamma^* \rightarrow q\bar{q} \rightarrow 2\pi$ will  
dominantly produce $2\pi$ systems of $J^P = 1^-$.  Indeed this is found when the $J^P = 1^- 
$ state is projected out.  This parton-hadron duality approach allows detailed predictions to be  
made for $\rho$ electroproduction. 
 
The predictions may be obtained by using the results of (\ref{eq:52})--(\ref{il}) to calculate  
the production of $u\bar{u}$ and $d\bar{d}$ pairs of mass $M$ via two gluon exchange.  We  
put $m_c = 0$ and change the integration variable in (\ref{eq:52}) and (\ref{fcct}) from  
$k_t$ to $\cos \theta$, where $\theta$ is the polar angle of the outgoing $q$ in the $q\bar{q}$  
rest frame with respect to the direction of the incoming proton.  Thus the transverse  
momentum of the outgoing quark is 
\be 
\label{eq:74d} 
k_t \; = \; \textstyle{\frac{1}{2}} M \: \sin \theta. 
\ee 
Then (\ref{eq:52}) and (\ref{fcct}) can be shown to give the $q\bar{q}$ production cross  
sections \cite{MRT}
\begin{eqnarray} 
\label{eq:74e} 
\fl \frac{d\sigma_L}{dM^2} & = & \frac{4 \pi^2 e_q^2 \alpha}{3 B_\rho} \:  
\frac{Q^2}{(Q^2 + M^2)^2} \: \frac{1}{8} \int_{-1}^1 d \cos \theta \: \left | d_{10}^1  
(\theta) \right |^2 \: |I_L|^2, \\ 
\label{eq:74f} 
\fl \frac{d \sigma_T}{dM^2} & = & \frac{4 \pi^2 e_q^2 \alpha}{3 B_\rho} \:  
\frac{M^2}{(Q^2 + M^2)^2} \: \frac{1}{4} \int_{-1}^1 d \cos \theta \left ( \left |d_{11}^1  
(\theta) \right |^2 + \left |d_{1-1}^1 (\theta) \right |^2 \right ) \: |I_T|^2, 
\end{eqnarray} 
where $d_{\lambda \mu}^J (\theta)$ are the usual rotation matrices.  The $d\ell_t^2$  
integrals $I_{T,L}$ over the transverse momenta $\pm \mbox{\boldmath $\ell$}_t$ of the  
exchanged gluons are given by (\ref{it}) and (\ref{il}).  They are essentially the cross sections  
for the interaction of the $(q\bar{q})_{L,T}$ pair with the proton.  If the $q\bar{q} + p$ cross section 
were constant (that is $I_L = I_T =$~constant) then we can perform the integrations in  
(\ref{eq:74e}) and (\ref{eq:74f}) and find $\sigma_L/\sigma_T = Q^2/4 M^2$.  An idea of  
the effects of the distortion of the $q\bar{q}$ state by the two-gluon exchange can be  
obtained by evaluating $I_T$ and $I_L$ of (\ref{it}) and (\ref{il}) in the leading $\log$  
approximation, in which it is assumed that the main contribution comes from the interval  
$\ell_t^2 < k_t^2$.  Then 
\be 
\label{eq:74g} 
I_L \; = \; I_T \; \sim \; (k_t^2)^\gamma / k_t^2 \; \sim \; (\sin^2 \theta)^{\gamma - 1}, 
\ee 
where $\gamma$ is the anomalous dimension of the gluon, see (\ref{eq:74b}).  If we  
substitute this behaviour in (\ref{eq:74e}) and (\ref{eq:74f}), project out the spin~1  
component and assume that $\gamma$ is constant over the region of integration then we  
obtain 
\be 
\label{eq:74h} 
\frac{\sigma_L}{\sigma_T} \; = \; \frac{Q^2}{M^2} \: \left ( \frac{\gamma}{\gamma + 1}  
\right )^2. 
\ee 
Now higher $Q^2$ means larger $x$ and both changes imply smaller $\gamma$.  As a consequence, 
the decrease of $\gamma$ with increasing $Q^2$ has the effect of masking the $Q^2$ growth of  
$\sigma_L/\sigma_T$.  The $J^P = 1^-$ projection integrals over $\cos \theta$ turn out to be linearly dependent on  
the $I_i$ (i.e.~not on $|I_i|^2$ as in (\ref{eq:74e}) and (\ref{eq:74f})) and are 
therefore less infrared sensitive than (\ref{eq:74e}) and (\ref{eq:74f}).   
$\sigma_T$, as well as $\sigma_L$, is convergent as $\theta  
\rightarrow 0$ provided only that $\gamma > 0$. 
 
The complete calculation does not, of course, make the above simplifications.  The latest  
predictions \cite{MRT4}, which are shown by the continuous curve in Fig.~\ref{rholt}, uses a  
skewed unintegrated gluon distribution, which is determined from a conventional gluon found  
in a global parton analysis.  Soft gluon emissions and the contribution of the real part of the  
amplitude\footnote{The optical theorem determines the imaginary part and a dispersion  
relation then provides an estimate of the real part.} are also included.  $\sigma_L$ is found to  
be insensitive to the infrared, whereas $\sigma_T$ is more sensitive but the uncertainty is less  
than the variation due to using gluons of different global analyses.  The absolute  
normalization, on the other hand, is again subject to uncertainties.  One is the higher order  
corrections which give rise to a substantial $K$ factor, the second is the size of the $\Delta  
M$ interval taken about the $\rho$ resonance.  Fortunately these ambiguities do not affect the  
ratio $\sigma_L/\sigma_T$.  Moreover the {\it shapes} of the $Q^2$ and $W$ distributions  
can be predicted reliably.  The enhancement effect of using the skewed  
distribution\footnote{An even larger enhancement, by a factor of about 2, comes from using  
skewed distributions to describe $\Upsilon$ photoproduction, $\gamma p \rightarrow  
\Upsilon p$ \cite{MRT3,MCD}.} increases with $Q^2$.  The QCD prediction for the $Q^2$  
shape agrees well with the HERA data \cite{MRT4}. 
 
So far the discussions have assumed $s$ channel helicity conservation (SCHC) which means  
that the produced $\rho$ meson retains the helicity of the incoming virtual photon.  However  
a similar QCD model to that discussed above, predicts a small violation of SCHC given by the 
helicity-flip amplitude describing the $\gamma_T \rightarrow \rho_L$ transition, which satisfies \cite{IvKi} 
\be 
\label{eq:74i} 
\frac{{\rm helicity\mbox{-}flip~amplitude}}{{\rm helicity\mbox{-}conserving~amplitude}}  
\; = \;  
\frac{\sqrt{|t| / 2}}{Q \; \gamma}. 
\ee 
The correlations of the observed angular distributions of $\gamma^* \rightarrow \rho$  
production and the $\rho  
\rightarrow \pi^+ \pi^-$ decay allow 15 density matrix elements to be measured so that the  
consequences of (\ref{eq:74i}) can be tested and, indeed, the fully helicity density matrix can be explored 
\cite{WOLF}.  The QCD predictions \cite{IvKi,KNZ} are in excellent agreement with HERA data. 
      
\section{Diffractive $J/\psi$ production at large $|t|$ and quasi-elastic       
$\gamma p$ scattering}      
      
Diffractive $J/\psi$ production at large $|t|$ and even more so    
quasi-elastic $\gamma p$ scattering are special cases of two particle    
elastic scattering as depicted in Fig.~\ref{large_t} where the final    
states $A'$ and $B'$ might differ from the initial state particles    
$A$ and $B$. The momentum transfer $t$ is supposed to be large enough    
so that perturbation theory is applicable but still much smaller than    
the total energy $W$. As for inclusive and exclusive    
diffractive production at $t=0$ the driving mechanism is two gluon    
exchange at high energies where $W^2 \gg |t|$.     
\begin{figure}[htb]        
  \vspace*{0cm}       
\begin{center}      
\epsfig{figure=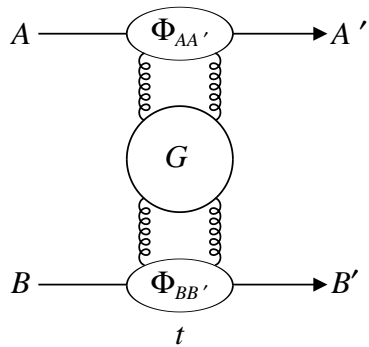,bbllx=4cm,bblly=15cm,bburx=13cm,bbury=24cm,width=6cm}  
\end{center}       
\vspace*{-2cm}       
\caption{\it Schematic representation of (\ref{eq:84a}).  $\Phi_{AA^\prime}$ is the impact   
factor for the $A \rightarrow A^\prime$ transition via two gluon exchange, while $G$ is the   
Green function for the two interacting (Reggeized) gluons.  \label{large_t}}    
\end{figure}    
Unlike the previous subsection where we had a two-scale problem with    
a hard scale at the top ($Q^2$) and a soft scale at the bottom we now have    
a single scale problem ($t$). In this domain contributions of the form    
$\alpha_S \log(W^2/|t|)$ are dominant rather than $\alpha_S \log(Q^2/Q_0^2)$,    
and they must be resummed to all orders. This is accomplished by the     
non-forward BFKL equation \cite{Lip}. Let us sketch the application to    
elastic or diffractive scattering. The general structure of the amplitude for a process    
$A+B\to A'+B'$ is    
\be     
\label{eq:84a}    
A (A + B \to A' + B') \;=\;\Phi_{AA'} \otimes G \otimes \Phi_{BB'},    
\ee    
as sketched in Fig.~\ref{large_t}. The impact factors $\Phi_{AA'}$    
and $\Phi_{BB'}$ describe the transition $A\to A'$ and $B\to B'$,    
whereas $G$ is a universal Green function for two interacting gluons.     
The Green function, which we have previously called gluon ladder,    
satisfies the BFKL equation which has been derived and solved    
in the leading log approximation, where the leading     
$(\alpha_S\ln(W^2/|t|))^n$ terms are resummed.  In the next-to-leading $\log$     
approximation $\alpha_S (\alpha_S \ln (W^2/| t | )^n$ terms would also have to be resummed.  
So far NLO corrections have only been performed for $t = 0$ \cite{FL,CC,FF}.
    
In the case of diffractive vector meson production     
$A$ represents a photon (we will assume it     
to be real) and $A'$ a vector meson.     
The problem of modelling     
the vector meson wave function is still one of the main obstacles.     
Since we are interested    
in the dynamcs of the 'hard' exchange rather than the final state,      
it has been suggested to replace the vector meson by a photon    
\cite{IvWu,GiIv}. Then $A'$ is a photon again.     
The upper part of the diagram in Fig.~\ref{large_t} which describes the coupling of    
the two gluons to the photon has the same form as     
in (\ref{eq:A}) except that we have to incorporate non-zero momentum transfer. The     
amplitude for elastic $q\bar{q}$-proton scattering, $A_{q\bar{q}+p}$ in (\ref{eq:A}),     
contains the Green function $G$ and the lower impact factor $\Phi_{BB'}$.     
In most cases the impact factor is defined in momentum space but for reasons which become    
clearer later on we work in impact parameter space.    
    
The amplitudes describing\footnote{The reason why it is sufficient to consider the scattering off a quark, 
rather than a proton, is given above (\ref{eq:89a}).} $\gamma q \rightarrow \psi q$ and $\gamma q \rightarrow \gamma     
q$ at large $|t|$ are of the form of (\ref{eq:84a}) \cite{IvWu}
\be    
\label{eq:85a}    
\fl A (\gamma + q \rightarrow (\psi, \gamma) + q) \: \sim \: \int d\nu \frac{\nu^2}{(\nu^2 +   
1/4)^2} \: \left ( \frac{\hat{s}}{|t| + M_V^2} \right )^{\omega (\nu)} \: \Phi_{\gamma (\psi,     
\gamma)}^\nu \: \Phi_{qq}^{\nu *}    
\ee    
where $M_V = 0, M_\psi$ and $\sqrt{\hat{s}}$ denotes the total energy at the parton level,     
since the exchange gluons couple to partons rather than to the proton as a whole.  $\omega$ is     
the \lq famous\rq~BFKL exponent \cite{BFKL}    
\be    
\label{eq:86a}    
\omega (\nu) \; = \; \frac{3 \alpha_S}{\pi} \: \left [ 2 \Psi (1) \: - \: \Psi \left (     
\textstyle{\frac{1}{2}} + i \nu \right ) \: - \: \Psi \left ( \textstyle{\frac{1}{2}} - i \nu \right )     
\right ]    
\ee    
where $\Psi$ is the logarithmic derivative of the Euler gamma function, $\Psi (z) \equiv     
\Gamma^\prime (z)/\Gamma (z)$.  It is the key result from the resummation of the $(\alpha_S     
\log (\hat{s}/|t|))^n$ terms by the BFKL equation.  It controls the power behaviour with   
regard to $\hat{s}$, that is the energy dependence of the process.  This is why we speak of the   
QCD or \lq hard\rq~Pomeron.  The general form for the \lq impact\rq~factor   
$\Phi_{AA^\prime}$ for the upper blob of Fig.~\ref{large_t} is\footnote{For photon  
production there are   
actually two different impact parameters $\bar{\Phi}_{\gamma\gamma}$.  One comes from   
$\psi^\gamma(\lambda_\gamma=\pm 1) \psi^{\gamma *}(\lambda_\gamma=\pm 1)$  
and one from the helicity flip product  
$\psi^\gamma(\lambda_\gamma=\pm 1) \psi^{\gamma *}(\lambda_\gamma=\mp 1)$.}  
\be    
\label{eq:87a}    
\fl \bar{\Phi}_{\gamma (\psi, \gamma)}^\nu \; \sim \; \int d^2 \mbox{\boldmath $r$}_1 d^2     
\mbox{\boldmath $r$}_2 \: \psi^\gamma (\mbox{\boldmath $r$}) \: E_\nu (\mbox{\boldmath     
$r$}_1, \mbox{\boldmath $r$}_2) \: \psi^{\psi, \gamma *} (\mbox{\boldmath $r$}) \: e^{i     
\alpha \mbox{\boldmath $q$} \cdot \mbox{\boldmath $r$}_1} \: e^{i (1 - \alpha)     
\mbox{\boldmath $q$} \cdot \mbox{\boldmath $r$}_2}    
\ee    
with $\mbox{\boldmath $r$} = \mbox{\boldmath $r$}_1 - \mbox{\boldmath $r$}_2$ and     
$\mbox{\boldmath $q$}^2 = |t|$ and where $\psi (\mbox{\boldmath $r$})$ is given by (\ref{ftwf}).  
Since now $|t| \neq 0$ we have to introduce two impact     
parameter variables $\mbox{\boldmath $r$}_1, \mbox{\boldmath $r$}_2$, one for each of     
the quark and the antiquark produced in the $\gamma \rightarrow q\bar{q}$ transition with longitudinal 
momentum fractions $\alpha$ and $1-\alpha$ respectively.  An integration over $\alpha$ is implied.       
$\bar{\Phi}$ is not quite the impact factor $\Phi_{AA^\prime}$ of (\ref{eq:84a}).  First, part   
of the     
Green function is included by the eigenfunction    
\be    
\label{eq:88a}    
E^\nu (\mbox{\boldmath $r$}_1, \mbox{\boldmath $r$}_2) \; = \; \left ( \frac{r^2}{r_1^2     
r_2^2} \right )^{\frac{1}{2} + i \nu}    
\ee    
of the non-forward BFKL equation and, second, it involves the convolution ($\otimes$) over     
the impact parameter variables $\mbox{\boldmath $r$}_1, \mbox{\boldmath $r$}_2$.  The     
effect of the net transverse momentum $\mbox{\boldmath $q$}$ flowing in the $t$ channel is     
embedded in the phase factors in (\ref{eq:87a}).  The wave functions $\psi^\gamma     
(\mbox{\boldmath $r$})$ and $\psi^\psi (\mbox{\boldmath $r$})$ of the photon and $J/\psi$     
are given, respectively, by (\ref{impact-t-wave}) and the \lq massive\rq~version of   
(\ref{impact-l-wave}).  Note that after taking the Fourier transform of $\Psi^\psi (\mbox{\boldmath   
$k$}_t)$ of (\ref{psiv}) with respect to $\mbox{\boldmath $k$}_t$, we are left with $\delta (\alpha   
- 1/2)$.    
    
Large $|t|$ has the advantage that factorization works on the proton side. The       
Pomeron couples to individual partons rather than the whole proton.       
This statement is not as obvious as it seems for an exchange       
in the form of BFKL-ladder  diagrams. It has been      
proven, though, in Refs.~\cite{ForRys,MueTan,BFLLRW}.       
The impact factor with respect to a quark-line reads:      
\beqn      
\label{eq:89a}    
\bar{\Phi}_{qq}^\nu &\sim& 2\;\int d^2\mbox{\boldmath $r$}_1\;      
\left(\frac{1}{\mbox{\boldmath $r$}_1^2}\right)^{1/2 + i \nu}\;e^{i \mbox{\boldmath $q$}     
\cdot \mbox{\boldmath $r$}_1}.      
\eeqn      
The factor 2 in front reflects the fact that either of the two gluons      
in the ladder diagram couple to the quark. For a gluon line      
the difference is only the colour factor.      
      
Using the amplitudes (\ref{eq:85a}) we may calculate $d\sigma/dt$ for both $\gamma q     
\rightarrow \psi q$ and $\gamma q \rightarrow \gamma q$ at large $|t|$.  The major work goes     
in evaluating the integrals in the impact factors (\ref{eq:87a}) and (\ref{eq:89a}).  The     
numerical analysis of both cross sections shows a strong enhancement due to       
the BFKL resummation of leading $(\alpha_S \log (\hat{s}/|t|))^n$ terms. The analysis also      
shows that both cross sections are of the same order of magnitude and very rapidly      
rising with increasing energy. The expected      
statistics at HERA should allow an experimental study of these processes.       
An important issue are NLO corrections to the BFKL-kernel \cite{FL,CC}. For $t=0$ they were      
found to be very large and negative. For $t \ne 0$, however, these corrections      
are not yet known. Nevertheless the general expectation is that they will reduce the theoretical      
prediction for the cross section.      
The quasi elastic photon-quark scattering can also be studied with virtual photons      
in the initial state. This would reintroduce the scale $Q^2$ \cite{EvFor}.  Once      
the impact factors are calculated they can be recombined in various ways. One example       
would be elastic $\gamma \gamma$-scattering at high      
energies. The limit $t \to 0$ can be studied in the case of       
$\gamma^* \gamma^* \to \gamma \gamma$. Deeply Virtual Compton Scattering (DVCS) \cite{FFS},       
i.e. the process $\gamma^* p \to \gamma p$, is the analog process in deep inelastic scattering.      
The limit $t \to 0$ as such is perturbatively defined, the $t$ dependence is sensitive      
to non-perturbative effects.

\section{Non-factorization in diffraction at hadron colliders \label{sect-hadron}}      
      
Historically diffractive scattering was measured in hadron collisions first.      
In the case of single inclusive diffraction one of the hadrons is       
tagged in the detector whereas the other dissociates. This process itself is, of course,      
purely soft, but with increasing energies the opportunity has emerged of measuring      
jets in the final state or similar hard processes (such as heavy flavour or $W$-boson     
diffractive production).      
      
The first idea of how to interpret hard diffractive processes       
was to view the Pomeron as a quasi-real hadron with      
a parton structure \cite{IngSchl} that allows the description of hard scattering      
in analogy to usual hadron-hadron scattering. One only needs to define      
parton distributions for the Pomeron, to be determined by experiment,      
and a Pomeron flux-factor. The latter represents the probability for      
a hadron to radiate off a Pomeron. This probability, of course, should be      
independent of the hard subprocess and consistent with soft inclusive      
scattering. One can assume the following factorization of the soft      
diffractive cross section into a flux factor $f_\funp$      
and the hadron-Pomeron cross section $\sigma^{\funp}$:      
\beqn\label{regge-fact}      
\frac{d \sigma^D}{dt d x_{\fP}}&=&f_{\fP}(x_{\fP},t)\;\sigma^{\fP}(M^2), \\      
f_{\fP}(x_{\fP},t)&=&\frac{\beta_{\fP h}^2 (t)}{16 \pi}\;      
\left(\frac{s}{M^2}\right)^{2\alpha_{\fP}-1}\;\;.      
\eeqn      
The variable $x_{\fP}$ has the same meaning as in previous sections (i.e. $x_{\fP} \approx  
M^2/s$ at small $Q^2$) and $\beta_{\fP h}$ describes the coupling of the Pomeron to a  
hadron, see  
Fig.~\ref{hard_diff}.      
The normalization of the flux-factor $f_\funp$ is unfortunately ambiguous,       
\begin{figure}[htb] 
\vspace*{-0cm} 
\begin{center}
 \epsfig{figure=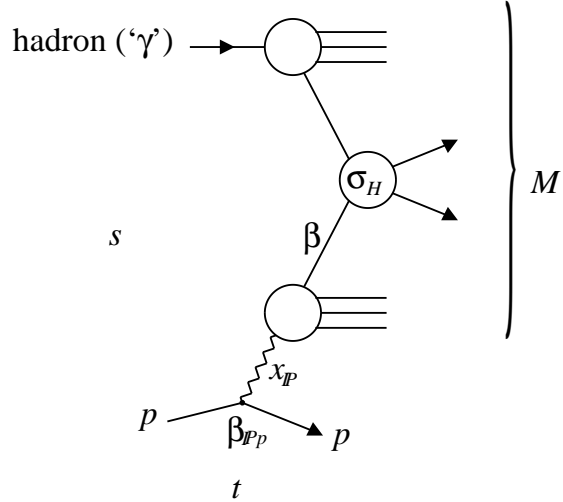,height=7cm} 
\end{center}
\vspace*{-0.8cm} 
\caption{\it Schematic diagram of diffractive hadronic scattering via a partonic constituent 
of the Pomeron which assumes that the structure of the upper hadron can be factored off.   
$\sigma_H$ denotes the cross section of the \lq hard\rq~partonic subprocess.  Unfortunately  
this partonic picture is not valid, since interactions with the remnants of the  
hadron 
cannot be neglected. \label{hard_diff}} 
\end{figure} 
since the Pomeron is not a real particle. This ambiguity poses       
serious problems in defining the correct normalization for the prediction      
of any hard cross section. According to section \ref{fact}       
we can define diffractive parton distributions for       
deep inelastic scattering. After dividing out the flux-factor one should arrive       
at a definition for the parton distribution of the Pomeron:       
\beqn      
\label{eq:a102}    
q^{\fP}(\beta)&=&\frac{q^D(\beta,x_{\fP},t)}{f_{\fP}(x_{\fP},t)}\\      
\label{eq:a103}    
g^{\fP}(\beta)&=&\frac{g^D(\beta,x_{\fP},t)}{f_{\fP}(x_{\fP},t)}.    
\eeqn      
For real hadrons the momentum sum of all partons      
adds up to the total momentum of the hadron. This momentum sum rule provides an     
important normalization constraint on the determination of the parton distributions.  Here     
there is no such constraint; (\ref{eq:a102}) and (\ref{eq:a103}) only fix the relative quark     
versus gluon content in the Pomeron.  Another problem arose when the predictons for   
diffractive processes were extrapolated from the CERN SPS-collider      
energy $(\sqrt{s} = 630$~GeV) to the Fermilab Tevatron proton-antiproton collider energy     
$(\sqrt{s} = 1800$~GeV). The predicted cross sections overshot the measured      
values significantly. The origin of this disagreement has never been completely resolved.     
Unitarity corrections provide a possible explanation \cite{GLM}.      
For data from the HERA electron-proton collider, a consistent treatment of diffractive deep     
inelastic scattering and diffractive photoproduction of jets in terms of the parton distributions     
of the Pomeron seems possible \cite{ColWhit} without      
renormalizing the flux-factor.  However predictions for hard diffractive processes      
at the Tevatron, based on Pomeron distributions extracted from HERA data, fail badly unless     
one assumes a renormalization of the Pomeron flux-factor \cite{Dino}.      
      
The problem is that although factorization has been proven for diffractive deep inelastic     
scattering, it is not valid in diffractive hadron scattering \cite{Col}.       
To discuss the issue  of non-factorization in diffractive hadron scattering let us suppose that a     
quasi-real photon is a model for a hadron, and make use of our previous formalism.      
When we talk about factorization we have to specify three different cases:      
first, Regge-type factorization (\ref{regge-fact}), second, collinear factorization of the     
Pomeron and third collinear factorization      
of the photon (hadron). We have already mentioned problems related      
to Regge-type factorization due to the ambiguity in the definition of the       
Pomeron flux-factor. In the genuine QCD-approach the Pomeron is solely viewed      
as a $t$ channel vacuum exchange, and is not associated with real particles.      
The question whether or not a Pomeron trajectory exists, which passes through physical     
(glue-ball) states for positive values of $t$, remains open.       
      
A closer look at the diagrams presented in Fig.~5 shows the origin of the   
violation of collinear factorization of the structure of the quasi-real photon, which we use as a   
model for the structure of the hadron. For example let us look at the triple Regge domain where we can 
assume a large transverse momentum      
in the final state while keeping the virtuality of the photon small and hadron-like.       
\begin{figure}[htb]       
\begin{picture}(15,5)(1,1)      
\put(0,0){\epsfig{figure=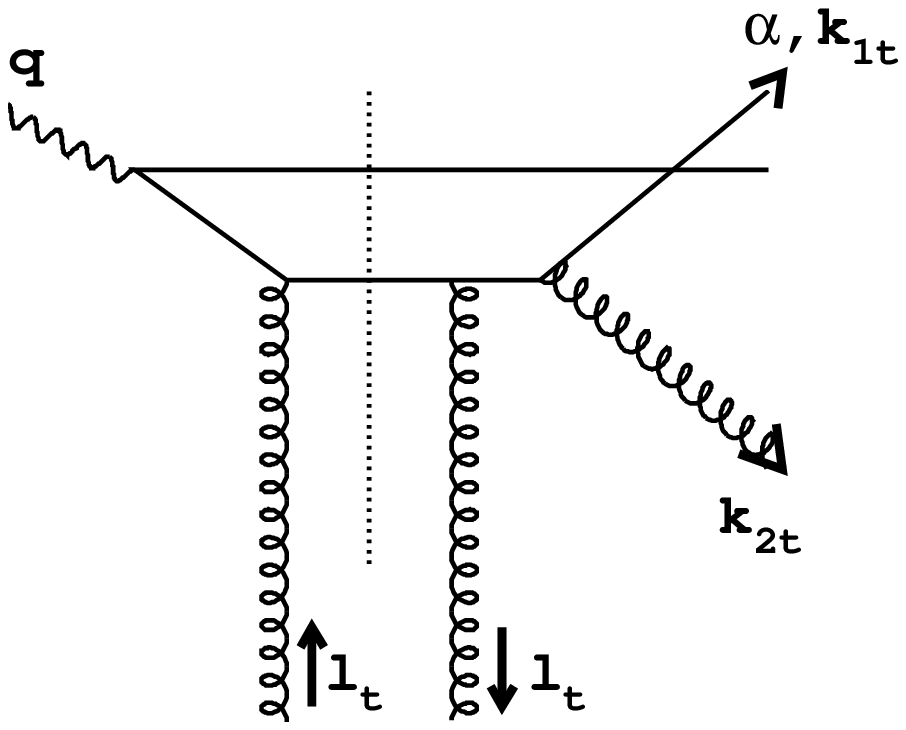,height=5.5cm}}      
\put(7,3){\begin{Huge}$\rightarrow$ \end{Huge}}      
\put(8,0){\epsfig{figure=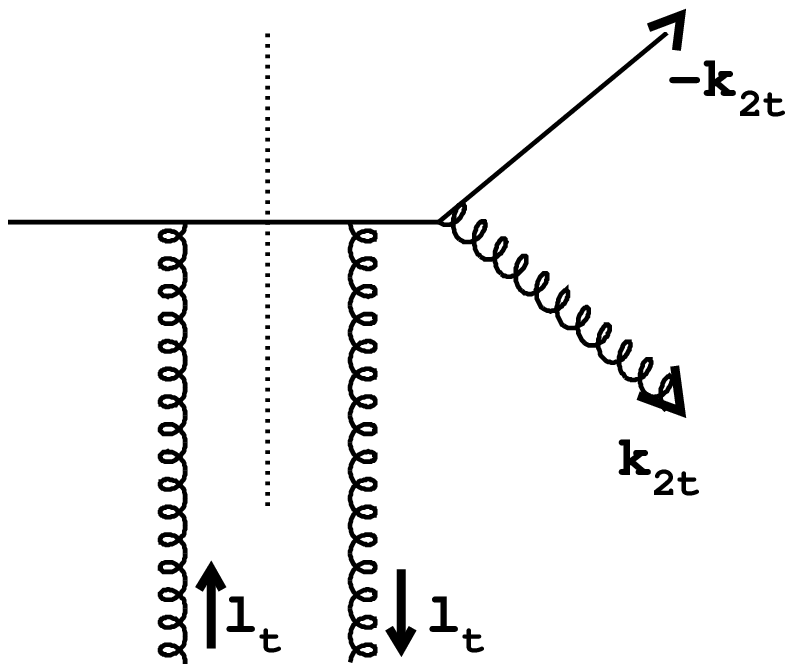,height=5.5cm}}      
\end{picture}      
\caption{\it The first diagram is an example of the subset of diffractive $q\bar{q}g$   
production diagrams in which the $t$ channel gluons couple to a single quark.  A quasi-real   
photon is used to mimic an incoming hadron.  It appears possible to factor off the \lq   
photon\rq~structure and to use the second diagram to describe the diffractive hadroproduction   
of two jets by means of the hard subprocess $q + p \rightarrow (qg) + p$.  The reason why  
this, in fact, is not possible is explained in the text.  
\label{diag8}}      
\end{figure}      
The left diagram of Fig.~\ref{diag8} is one out of the complete set of diagrams      
which contribute to expression (\ref{fqqgtr}) for diffractive $q\bar{q}g$ production.  The   
particular feature is that      
only one quark line is involved in the scattering process, while the second has       
no interaction with a gluon. This second quark plays the role of the       
hadron-remnant. If we add up all related diagrams which leave the remnant quark       
unscattered, the gluon exchange amplitude simplifies from (\ref{mil}) to    
\beqn \label{til2}      
\fl \bar{M}_{ij} &=& \left(\frac{\mbox{\boldmath $k$}_1 + \mbox{\boldmath     
$k$}_2}{D(\mbox{\boldmath       
$k$}_1 + \mbox{\boldmath $k$}_2)} \right)_i\; 
 \int \frac{d^2 \mbox{\boldmath $\ell$}}{\pi \ell^2} \: \alpha_S {\cal  
F}     
(x_\funp, \ell^2) \:       
     \left( \frac{\mbox{\boldmath $k$}_2+\mbox{\boldmath $\ell$}}{(\mbox{\boldmath       
$k$}_2+\mbox{\boldmath $\ell$})^2}+ \frac{\mbox{\boldmath $k$}_2-\mbox{\boldmath       
$\ell$}}{(\mbox{\boldmath $k$}_2-\mbox{\boldmath $\ell$})^2}        
                  - 2\;\frac{\mbox{\boldmath $k$}_2}{\mbox{\boldmath $k$}_2^2} \right)_j,  
\eeqn      
where $i, j = 1, 2$ are the vector components in the transverse plane.  The first term, which  
we denote $a_i \equiv \left ( \cdots \right )_i$, involves the quark propagator, and the second  
term $b_j \equiv \left ( \cdots \right )_j$ which is associated with the $s$   
channel gluon, is that shown in     
(\ref{mil})  together with the $\mbox{\boldmath $\ell$} \rightarrow -\mbox{\boldmath  
$\ell$}$ term.  The dominant configuration is $\mbox{\boldmath $k$}_1 \approx -  
\mbox{\boldmath $k$}_2$, describing a quark and gluon jet in the final state, such that  
$\mbox{\boldmath $k$}_2^2 \approx \mbox{\boldmath $k$}_1^2 \gg (k_1 + k_2)^2 \sim  
Q^2$.  The factorization in this case follows since    
\be    
\label{eq:a105}    
\bar{M}_{ij} \: \bar{M}_{ij}^* \; \sim \; (a_1^2 + a_2^2) (b_1^2 + b_2^2).    
\ee    
The first factor, originating from the $a_i \equiv \left ( \cdots \right )_i$ in (\ref{til2}), leads     
to a logarithm in $\mbox{\boldmath $k$}_2^2/Q^2$      
in the cross section. This factor would be divergent for $Q^2=0$. It      
represents a collinear singularity which under usual circumstances      
is absorbed into the structure of the photon.      
      
It appears that exactly the same formalism is appropriate to describe diffractive hadron     
scattering with jets in the final state.  It seems that we simply need to calculate the process     
$q+p \to qg+p$ given by the second diagram of Fig.~\ref{diag8}.      
In the triple Regge limit, which in this case means  $M^2\gg \mbox{\boldmath $k$}_2^2$,      
one finds the same structure for the $s$ channel gluon contribution as in the amplitude      
(\ref{til2}). This is of course expected due to the factorization      
of the photon structure mentioned earlier. By integrating over the azimuthal      
angle of $\mbox{\boldmath $\ell$}$, after some algebra, we find:      
\beqn\label{theta}      
\fl \int_0^{2\pi}\frac{d\varphi_l}{2\pi}\;       
\left(\frac{\mbox{\boldmath $k$}_2+\mbox{\boldmath $\ell$}}{(\mbox{\boldmath       
$k$}_2+\mbox{\boldmath $\ell$})^2}+ \frac{\mbox{\boldmath $k$}_2-\mbox{\boldmath       
$\ell$}}{(\mbox{\boldmath $k$}_2-\mbox{\boldmath $\ell$})^2}        
                  - 2\;\frac{\mbox{\boldmath $k$}_2}{\mbox{\boldmath $k$}_2^2}\right)\;=\;      
-2\;\Theta(\ell^2 - k_2^2)\;\frac{\mbox{\boldmath       
$k$}_2}{\mbox{\boldmath $k$}_2^2}\;\;.      
\eeqn      
That is the low $\ell$ contribution cancels out such that the integration     
over $\ell^2$ has a lower hard cutoff      
given by $k_2^2$. The process $q+p \to qg+p$ thus seems to be   
calculable completely      
pertubatively. An identical conclusion has recently been made      
in Ref.~\cite{YuanChao} where the same process has been calculated      
without restricting it to the triple Regge limit.       
\begin{figure}[htb]       
\begin{center}\vspace*{-0.5cm}      
\epsfig{figure=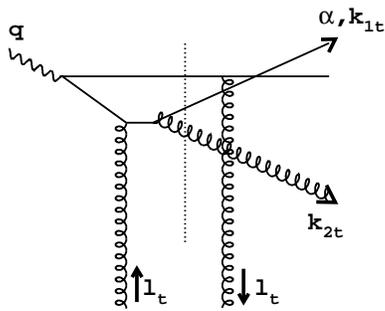,height=5.5cm}      
\end{center}\vspace*{-1.5cm}      
\caption{\it A contribution involving the scattering of the remnant which not only ruins the   
possibility of using parton distributions of the Pomeron to describe hadronic diffractive   
scattering, but, even more, prevents us factoring off the structure of the quasi-real photon that   
is mimicking an incoming hadron.  \label{diag9}}      
\end{figure}      
The problem is that there is a non-negligible rescattering of the partons with spectator partons 
due to \lq soft\rq~gluon exchange.  Free quarks or free      
gluons do not exist. Quarks or gluons can only be part of the initial state for those processes     
where the factorization theorem is applicable.      
The true initial state is colourless (hadrons, photons, etc.) and single partons      
are always accompanied by a remnant which eventually       
takes part in the interaction.      
This is precisely what happens in the case of diffraction and is the reason why       
factorization fails for diffractive hadron scattering \cite{SOPER}. One of the      
$t$ channel gluons interacts with the remnant, as depicted in Fig.~\ref{diag9},      
and upsets the subtle cancellation that leads to the result (\ref{theta}).       
Taking into account all possible interactions which involve the remnant brings us back      
to the complete expressions (\ref{fqqgtr}) and (\ref{mil}) of Section \ref{triple-regge}.      
    
Insight into the problem may be obtained from a simple estimate of the cross section.  To     
obtain the diffractive hadronic cross section we have to integrate $\ell^2$ from $Q_0^2$ up     
to $\mbox{\boldmath $k$}_1^2 \approx \mbox{\boldmath $k$}_2^2$, where $Q_0$ denotes   
the typical hadronic scale.  The dominant     
contribution comes from the lower end of the integration so we may put $\ell^2 \ll k_2^2$ in     
(\ref{mil}) and obtain    
\be    
\label{eq:a107}    
M_{ij} \; \sim \; \int_{Q_0^2} \frac{d^2 \mbox{\boldmath $\ell$}}{\ell^2} \: \alpha_S \: {\cal   
F} (x_\funp, \ell^2) \: \frac{1}{k_2^2} \left ( \delta_{ij} \: - \: \frac{(\mbox{\boldmath   
$k$}_2)_i (\mbox{\boldmath $k$}_2)_j}{k_2^2} \right ).    
\ee    
If we further assume ${\cal F} \sim 1/\ell^2$ then we find the diffractive cross   
section\footnote{The $\mbox{\boldmath $k$}_1^2$ integration over the interval   
$(\mbox{\boldmath $k$}_1 + \mbox{\boldmath $k$}_2)^2 \lapproxeq Q_0^2$ brings an   
extra factor of $Q_0^2$ which partially cancels the apparent $1/Q_0^4$ behaviour which  
follows from (\ref{eq:a107}).}  
\beqn      
\label{eq:a99}  
\frac{d\sigma^D}{dt dM^2 dk_2^2}&\sim&\frac{1}{M^2\;Q_0^2\;k_2^4}\;\;.      
\eeqn      
On the other hand if we were to use (\ref{theta}) to obtain the cross section from the   
perturbative subprocess $q+p\to qg+p $ then we would obtain      
\beqn      
\label{eq:a100}
\frac{d\sigma(q+p\to qg+p)}{dt dM^2 dk_2^2}&\sim&\frac{1}{M^2\;k_2^6},    
\eeqn      
which is power-suppressed by $Q_0^2/k_2^2$ relative to the true   
answer of (\ref{eq:a99})\footnote{A related discussion on diffractive $b\bar{b}$-production 
in hadron collisions can be found in \cite{Genya}.  However the results are not directly 
comparable to (\ref{eq:a99}) and (\ref{eq:a100}) since the soft interaction with the 
remnant in Fig.~3 of \cite{Genya} is neglected.}. This suppression      
is due to the cancellations in the angular integration       
of (\ref{theta}) and the consequent restriction of the integration      
over $\ell^2$. After connecting a gluon to the remnant, on the other      
hand, the integration over $\ell^2$ ranges down to the soft scale       
$Q_0^2$ boosting the cross section by the factor $k_2^2/Q_0^2$.       
We conclude that we must use the full non-factorizing formula (\ref{fqqgtr}).  This example   
clearly demonstrates that the well known factorization of the cross section into universal  
parton distributions      
and a \lq hard\rq~partonic subprocess is violated for colourless two-gluon exchange.      
The same conclusion holds for multiple gluon exchange. These findings are      
of course unfortunate with regard to practical applications, as they do not      
provide any quantitative description of hard diffraction in hadron collisions since we need to   
know the hadronic analogue of the virtual photon wave function.      
It offers, however, a qualitative explanation why the physics      
of diffraction in deep inelastic scattering is fundamentally different      
from diffraction in hadron scattering. The best way to proceed towards      
a quantitative description of diffractive hadronic processes is to model the wave function of  
the incoming hadrons.      
      
Issues related to non-factorization are       
the rapidity gap survival probability \cite{Bjgap,GLMgap}       
in double  diffractive dissocation, i.e. both hadrons dissociate, and      
unitarity corrections. Multiple gluon exchanges lead to       
absorptive corrections which limit the growth of the cross section      
with increasing energy. Once      
there are a lot of $t$ channel gluon exchanges the probability of      
radiation into the gap is, of course, high as well. These soft raditions also       
reduce the fraction of  high-$p_t$ jets with a rapidity gap in between       
them \cite{MueTan}.      
        
\section{Conclusions}      
If we assume that QCD is indeed the theory of strong interaction then the Pomeron    
itself is to be constructed from QCD. The simplest model to produce a colourless    
exchange is two gluon exchange    
and we have spun our discussion of diffractive processes around this model.    
It provides a good starting point for further sophistications such as gluon ladders    
and multiple gluon exchange. Many features in diffractive deep inelastic scattering    
such as the $\beta$-spectrum and the $Q^2$-dependence directly    
follow from the forms of the wave functions of the photon. In lowest order in perturbation    
theory the photon dissociates into a quark-antiquark pair ($q\bar{q}$). The corresponding    
wave function is mainly built from the photon coupling to the quarks     
and the energy denominator of the virtual $q\bar{q}$ state. At higher order    
the wave function formalism can be extended to also include a gluon (that is the $q\bar{q}g$    
state).  Gluon radiation becomes important and even the dominant contribution for large     
diffractive masses $M$. Longitudinally polarized photons play a special role    
as they dominate the large $\beta$-regime despite the fact that these contributions    
are of higher twist nature. The remaining regime of medium $\beta$ is filled    
with the contribution from transversely produced $q\bar{q}$-pairs as illustrated in    
Fig.~\ref{ddrev_plot}. We have also discussed the triple Regge limit where the mass $M$ of  
the  diffractively produced system is much bigger than $Q$. This regime is interesting as it  
allows us to interpret the $q\bar{q}g$ final state as a system of two quark dipoles (the gluon is decomposed    
in colour space into a quark-antiquark pair) scattering on the proton.    
In addition it is of use in the analysis of diffractive 3-jet data.    
    
We have also discussed the issue of nonperturbative effects in open diffractive    
deep inelastic scattering. Multiple gluon exchanges have to be considered as corrections.    
It has been argued that these corrections only slightly affect the $\beta$-distribution    
as the photon wave functions do not change. Up to colour factors multiple gluon exchange    
can be recast into the unintegrated gluon distribution.     
    
The photon wave function formalism can be straightforwardly carried across to vector     
meson production. Light vector mesons show a fairly flat ratio of the longitudinal    
versus transverse cross section at large $Q^2$ which is a consequence    
of $\gamma$ decreasing with $Q^2$, see (\ref{eq:74h}).  The large $Q^2$ scale, together with the projection    
on an exclusive final state, allows perturbative evolution to take place inside    
the Pomeron, i.e. the Pomeron can be associated with perturbative gluon ladders    
(the so-called hard or QCD Pomeron). As a consequence the energy dependence is much    
steeper than expected from a soft Pomeron. A similar conclusion can    
be drawn for $J/\psi$ photoproduction and open charm production.  In this case the hard scale    
is given by the charm mass.  Precise data are becoming available at HERA for these  
processes.  So   
far the measurements show evidence of the QCD expectations.  Another closely related  
process is the photoproduction of photons or vector mesons at large momentum transfer $|t|$.  
In this scenario the large scale is provided by $t$, but unlike deep inelastic scattering $t$  
stretches along the whole gluon ladder. Therefore a different evolution scheme following the  
BFKL equation is at work.    
    
We have demonstrated that diffractive deep inelastic scattering is consistent with collinear 
factorization and we have introduced     
diffractive parton distributions accordingly. On the other hand we have argued    
against general parton distributions of the Pomeron because factorization    
fails in diffractive hadron-hadron collisions. In particular we have shown    
the importance of the remnant of the dissociating hadron in estimating     
the diffractive cross section.    
    
In summary, the variety of diffractive processes in DIS, and the number of different kinematic   
distributions that are accessible to experiment, offer a sensitive probe of QCD.  The optical   
theorem means that the diffractive {\it amplitude} is proportional to the inclusive cross   
section.  The dependence on the {\it square} of a quantity magnifies the sensitivity of the   
diffractive probe.  Examples are $\gamma^* p \rightarrow Vp$ which depends on the square   
of the (skewed) gluon distribution of the proton, and $\gamma p \rightarrow J/\psi p$ at large   
$|t|$ which magnifies the BFKL-type enhancement effect.  Although much theoretical work   
remains to be done (such as the computation of higher order corrections), it is already clear   
that the accumulation of precise data for DIS diffractive processes will greatly deepen our   
understanding of QCD.  
      
\section{Acknowledgements}    
    
We thank Markus Diehl, Misha Ryskin and Thomas Teubner for reading through the  
manuscript and for     
numerous helpful comments.  One of us (MW) thanks the UK Particle Physics and     
Astronomy Research Council for support.

\end{document}